\documentclass[11pt]{article}
\newcommand{\Comment}[1]{{}}
\usepackage[textwidth = 430 pt, textheight = 630 pt]{geometry}
\usepackage{amssymb,euscript,amsmath,amsfonts}

\usepackage{color}
\definecolor{MyDarkBlue}{rgb}{0.15,0.15,0.45}
\usepackage[linktocpage=true]{hyperref}
\hypersetup{
colorlinks=true,
citecolor=MyDarkBlue,
linkcolor=MyDarkBlue,
urlcolor=MyDarkBlue,
pdfauthor={Chris Hull, Neil Lambert},
pdftitle={Time},
pdfsubject={hep-th}
}

\usepackage[numbers,sort&compress]{natbib}
\usepackage{hypernat}

\newcommand\ignore[1]{}
\def\one{{\,\hbox{1\kern-.8mm l}}}

\newcommand{\tr}{\operatorname{tr}}

\newcommand{\Cset}{{\,\,{{{^{_{\pmb{\mid}}}}\kern-.45em{\mathrm C}}}}}

\newcommand{\ie}{{\it i.e.~}}

\newcommand{\be}{\begin{equation}}
\newcommand{\ee}{\end{equation}}
\newcommand{\bea}{\begin{eqnarray}}
\newcommand{\eea}{\end{eqnarray}}

\newcommand{\Z}{\bf Z}

\newcommand{\R}{\mathbb R}
\begin{document}

 \renewcommand{\thefootnote}{\fnsymbol{footnote}}

\rightline{Imperial/TP/2014/CMH/01}
\rightline{CERN-PH-TH/2013-019}
\rightline{KCL-MTH-14-04}

   \vspace{1.8truecm}

\vspace{15pt}

\centerline{\LARGE \bf {\sc Emergent Time and the M5-Brane}   } \vspace{2truecm}

\centerline{
    {\large {\bf{\sc C.M.~Hull${}^{\,a,}$}}}\footnote{E-mail address:
                                 \href{mailto:Chris Hull <c.hull@imperial.ac.uk>}{\tt c.hull@imperial.ac.uk}}  {and}  \large {\bf {\sc N.~Lambert ${}^{\,b,c}$}}$^,$\footnote{E-mail address: \href{mailto:neil.lambert@kcl.ac.uk}{\tt neil.lambert@kcl.ac.uk}\\ \hbox{} \hskip0.375cm${}^b$Previous Address: 
                                 {\it Theory Division, CERN, 1211 Geneva 23, Switzerland}                                }}

\vspace{1cm}
\centerline{${}^a${\it Department of Physics}}
\centerline{{\it Imperial College London, UK}}

\vspace{.8cm}

\centerline{${}^b${\it Dept. of Mathematics}}
\centerline{{\it King's College London, UK}}
 \vspace{1.5truecm}

\thispagestyle{empty}

\centerline{\sc Abstract}
\begin{center}
\begin{minipage}[c]{380pt}{
 We consider the  maximal super-Yang-Mills theory  in 5 Euclidean  dimensions with   $SO(5)$ R-symmetry and 16 supersymmetries.   We  argue that the  strong coupling limit of this theory (with a possible UV completion) has an emergent time dimension and gives a description of the 5+1 dimensional Lorentz invariant $(2,0)$ theory of the M5-brane, compactified on a timelike circle with radius $R=g^2/4\pi^2$. 
  Our discussion involves issues of quantization of Euclidean theories without time.
  }

\end{minipage}
\end{center}

\vspace{.4truecm}

\noindent

\vspace{.5cm}

\setcounter{page}{0}

\newpage

\renewcommand{\thefootnote}{\arabic{footnote}}
\setcounter{footnote}{0}

\linespread{1.1}
\parskip 4pt

{}~
{}~

\makeatletter
\@addtoreset{equation}{section}
\makeatother
\renewcommand{\theequation}{\thesection.\arabic{equation}}

\section{ Introduction}

One of the more dramatic effects that can arise in a theory at strong coupling is the opening up of an extra spatial dimension, as in IIA string theory or 5-dimensional maximally supersymmetric Yang-Mills (SYM) theory\footnote{In this paper SYM always refers to the maximally supersymmetric case.}.
The aim of this paper is to investigate the possibility of an extra time dimension opening up in a similar way. Our main focus will be on SYM theory in 5 Euclidean dimensions with 16 supersymmetries, which we will argue has a strong coupling
 limit that is a theory in 5+1 dimensions, with a new time dimension opening up to give the (2,0) theory.  More generally, it is interesting to have a theory formulated with no time dimension but from which time emerges, and which  may be useful in thinking about cosmological models in which time and/or space are emergent. As we shall see, our discussion involves interesting questions about the role of time in quantum theory, and about the meaning of a compact time dimension. Our work also provides evidence for part of the network of dualities involving timelike
compactifications of \cite{Hull:1998vg, Hull:1998ym}, gives new insight into the (2,0) theory, and also makes contact with the (2,0) supersymmetric field theory constructed in \cite{lambert:2010wm}.

 The most famous example in which a new spatial dimension emerges is that of the IIA string, which at strong coupling becomes an eleven dimensional theory. The extra spatial dimension arises as a circle of radius $R=\sqrt{\alpha'}g_s$ where $g_s$ is the string coupling. The D0-branes of the IIA string become light at strong coupling and  are interpreted as Kaluza-Klein modes for M-theory compactified on a circle.
Another example is that of  SYM in 4+1 dimensions with 16 supersymmetries. The Yang-Mills coupling $g^2$ has dimensions of length and, at least for weak coupling, the theory is interpreted as the (2,0) theory  compactified on a circle of radius $R=g^2/4\pi^2$, with Yang-Mills solitons interpreted as Kaluza-Klein modes of the (2,0) theory.

There are a couple of issues arising with this. The first is that, although in both the type IIA and the 5-D
supersymmetric Yang-Mills theories
there is strong evidence that a new dimension arises at strong coupling, we do not yet have a fundamental formulation of M-theory or of the (2,0) theory. In the former case one can   regard M-theory as being defined by the strong coupling limit of type IIA string theory, including non-perturbative effects, whereas in the latter case the (2,0) theory is a well-defined conformal field theory 
 that arises as a decoupling limit of the M5-brane worldvolume theory \cite{Strominger:1995ac}, or of IIB string theory on K3 \cite{Witten:1995zh}.
 The M5-brane realisation
 leads one to identify the (2,0) theory as a holographic dual of M-theory on $AdS_7\times S^4$ \cite{Maldacena:1997re}. 
 The (2,0) theory  has also been formulated through deconstruction \cite{ArkaniHamed:2001ie} and as a  matrix-like model \cite{Aharony:1997th}. More recently there have also been  proposals on formulating 6D actions \cite{Chu:2012um} or 5D actions which include an explicit KK tower \cite{Ho:2011ni,Bonetti:2012st}. A definition of the (2,0) theory involving 5D SYM on ${\mathbb R}\times {\mathbb CP}^2$   with an integer coupling constant  has been suggested in \cite{Kim:2012tr}.

The second issue is that 5D SYM is not renormalisable, and has recently been shown to have an  UV divergence at 6 loops \cite{Bern:2012di}.
In discussing the quantum theory and its strong coupling limit, it is therefore necessary to embed SYM in a theory that is UV complete.
 It can be embedded in string theory as the worldvolume theory of a stack of D4-branes or as the heterotic string  compactified on a 5-torus; both of these have a 5D Yang-Mills sector, with
higher derivative corrections and couplings to other fields.
At strong coupling, the  D4-branes of the IIA string become a
stack of M5-branes wrapped on the M-theory circle so that the D4-brane worldvolume theory becomes the M5-brane worldvolume theory.
The heterotic string on ${\mathbb T}^5$ is dual to the IIB string on $K3\times S^1$ \cite{Hull:1994ys} and at strong coupling this circle decompactifies to give the IIB string on $K3$ \cite{Witten:1995ex}.
Thus in both constructions the strong coupling limit gives a theory containing the (2,0) field theory,   with the (2,0) filed theory  arising from the 5 dimensional  Yang-Mills sector. These constructions involve gravitational and other fields, but these are eliminated in the decoupling limit leading to the (2,0) superconformal field theory.

Of course, a natural UV completion of the 5D SYM would be the (2,0) theory on a circle  \cite{Seiberg:1997ax}, but it is the relation between the 5D SYM and the 6D (2,0) theory  at finite values of the radius that needs to be established.
On the other hand, if there were a way in which  5D SYM could be regarded as  defining a consistent quantum theory, then it  could be used to give a construction of the 6D (2,0) theory, in much the same way that M-theory can be defined as the strong coupling limit of type IIA string theory.
Indeed, there have been recent suggestions that the $4+1$ dimensional super-Yang-Mills theory could in fact be non-perturbatively well-defined with solitonic states providing a UV completion 
 \cite{Douglas:2010iu,Lambert:2010iw}. This need not be in contradiction with the perturbative divergences of the theory:
 a perturbative expansion in $g^2$ can be thought of as a low-energy expansion in the effective dimensionless coupling $g^2 E$, and this does not fully probe the high energy behaviour. Various relationships between this conjecture and the construction of \cite{ArkaniHamed:2001ie} \cite{Aharony:1997th} have been discussed in \cite{Lambert:2012qy}.

Dimensional reduction of the (2,0) theory on a spatial circle gives   SYM   in 4+1 dimensions.
Similarly, dimensional reduction of the (2,0) theory on a timelike circle  gives a SYM theory  in 5 Euclidean dimensions. We will argue that the quantization of this theory in 5 Euclidean dimensions gives a theory for which an extra time dimension opens up at strong coupling, so  that the theory is interpreted as the (2,0) theory compactified on a timelike circle of radius $R=g^2/4\pi^2$. As for the theory in 4+1 dimensions, the Euclidean SYM theory needs a UV completion, and we shall assume throughout that the theory we are discussing is embedded in a suitable UV completion, such as those discussed above.

A simple argument that this should be the correct quantum behaviour follows from the fact that SYM theory in 5 Euclidean dimensions is an analytic continuation of SYM in 4+1 dimensions. The path integral for the  theory in 4+1 dimensions is defined by Wick rotating to 5 Euclidean dimensions and a further analytic continuation to define a positive Euclidean action ($A_0 \to -i A_0$ {\it etc.}); the results of Euclidean calculations are then continued back to the theory in $4+1$ dimensions. The supersymmetric theory in 5 Euclidean dimensions arising from the timelike reduction of the (2,0) theory has an action which is not positive. Its quantization also involves an analytic continuation
to the {\it same} positive Euclidean action as for the $4+1 $ theory, as will be discussed in section 2.
Thus the same Euclidean functional integral governs both theories; the different physical results are obtained by different analytic continuations, one to the real supersymmetric  theory in 4+1 dimensions and one to the real supersymmetric theory in 5+0 dimensions.
In particular, the fact that the quantum theory features an extra dimension in one case should imply that it does in the other; the difference is that on continuing back to the real theory, in one case it is a space dimension, in the other it is a timelike one. This is of course rather formal, and one of the aims of this paper is to seek further evidence and physical intuition for this behaviour.

The (2,0) theory in 5+1 dimensions has $SO(5)$ R-symmetry and ADE gauge group G.
The timelike reduction gives a
 super Yang-Mills theory with 16 real supersymmetries in 5D Euclidean space with $SO(5)$
R-symmetry. A standard construction of SYM theories in $d$ dimensional Euclidean space is by dimensional reduction from $9+1$ dimensions on $9-d$ spatial dimensions and one time dimension. The resulting Euclidean  theory has non-compact R-symmetry $SO(9-d,1)$. For $d=4$, this gives a SYM theory in 5 Euclidean dimensions with R-symmetry $SO(4,1)$ and so this has the wrong R-symmetry to be the theory obtained from timelike reduction of the (2,0) theory.
The correct construction was identified in \cite{Hull:1999mt}.
In addition to the SYM theory in 9+1 dimensions, there is also one in  5+5 dimensions, as signature (5,5) also allows Majorana-Weyl fermions and the Fierz identities {\it etc.} are formally the same in these two signatures.
Dimensionally reducing on 5 timelike directions gives an SYM theory in 5 Euclidean dimensions
with $SO(5)$ R-symmetry and 16 supersymmetries, as required. This gives the direct construction of the Euclidean SYM theory we will be studying here.
The other SYM theory  in 5D Euclidean space with $SO(4,1)$ R-symmetry can be obtained from 5+1 dimensions from the timelike
reduction of the (1,1) supersymmetric SYM theory in 5+1 dimensions.

There are other situations in which a time dimension is expected to arise from a timeless theory.
In de Sitter holography \cite{Hull:1998vg}, \cite{Strominger:2001pn}, a conformal field theory in $D$ Euclidean dimensions is conjectured to be dual to a theory of gravity in $D+1$ dimensional de Sitter space in the limit of strong 't Hooft coupling, so that the holographic dual has an extra time dimension. In particular, 4D Euclidean SYM was conjectured to be dual to a string theory in 5 dimensional de Sitter space in \cite{Hull:1998vg}.
In \cite{ Hull:1998ym}, a version of string theory was proposed  in 10D Euclidean space, and it was argued that at strong coupling it gains an extra time dimension to become M-theory on a spacetime with a timelike circle.
As we shall see in section 8, this is closely related to the extra time dimension  arising here.
In the strong coupling limit of the Euclidean 5D SYM considered here, we expect the $SO(5)$ rotational symmetry to be enhanced to $SO(5,1)$ Lorentz symmetry.
This is reminiscent of the work of Horava on quantum gravity (for example see \cite{Horava:2009uw}) where Lorentz symmetry arises in the IR.


The rest of this paper is organized as follows. In section 2 we give the details of  Euclidean 5D SYM theory with $SO(5)$ R-symmetry.
In section 3 we discuss the 5D classical solutions arising from 4D YM instantons and their interpretation as modes  with discrete momentum arising from compactification of an extra dimension.
In section 4 we discuss a dynamical treatment of 5D SYM based on treating one of the spatial directions as a Euclidean time.   An examination of the the associated conserved charges leads to a  6D interpretation including an energy-momentum tensor and an extra timelike momentum. In section 5 we compare the BPS states of 5D SYM with the states of the 6D (2,0) theory   when treated using a similar notion of Euclidean time. In section 6 we look for evidence for the 6D SO(1,5) Lorentz symmetry in the 5D SYM. In section 7 we provide an alternative derivation of 5D SYM from the (2,0) system of equations of \cite{lambert:2010wm}. These lead to an identification of the charges of the (2,0) theory   arising in a standard hamiltonian treatment  in terms of quantities of 5D SYM. Section 8 relates the results here to M-theory, and in particular to the worldvolume theory of M5-branes wrapped on a compact time dimension. Finally section 9 gives a discussion of our results. We also include an appendix with details showing the infinite energy of singular string solitons.

We now comment on our notation. In this paper we will consider the 5+1 dimensional (2,0) theory as well as Euclidean 5D SYM. The former has an energy-momentum tensor which we denote by $T_{\mu\nu}$, $\mu,\nu = 0,1,..,5$ whereas the latter has a stress-tensor that we denote by $\Theta_{ij}$, $i,j =1,2,..,5$.  We will consider canonical formulations based on real time in the case 5+1 theory, but also canonical formulations in terms of a Euclidean time for both the 5+1 dimensional and 5D theories. Thus we encounter various charges that appear in the associated supersymmetry algebras. In particular we denote the standard 5+1 dimensional charges by  $P_\mu$, $Z^I_\mu$ {\it etc.}. When considering a Euclidean time formulation of this theory we denote the resulting charges by  $\hat P_\mu$, $\hat Z^I_\mu$ {\it etc}. For the 5D theory we denote the Euclidean charges by $\hat \Pi_0$, $\hat \Pi_i$, $\hat {\cal Z}^I_{i}$ {\it etc.}. In section 7 we introduce another set of charges $\tilde \Pi_0$, $\tilde \Pi_i $, $\tilde {\cal Z}^I_0$  {\it etc.} which are not {\it a priori} associated with a canonical formulation. Whilst these charges are all interrelated it is important to distinguish between them.

\section{ Euclidean 5-Dimensional Super-Yang-Mills}

The dimensional reduction of the (2,0) theory on time gives a Euclidean Yang-Mills theory with $SO(5)$ R-symmetry and 16 supersymmetries.
As discussed in the introduction, this theory can be constructed directly from the supersymmetric gauge theory in $5+5 $ dimensions by dimensionally reducing on the 5 timelike directions.
The 5+5 dimensional lagrangian is
\begin{align}
{\cal L}  =  \frac{1}{g^2} \tr \left[\frac{1}{4 }F_{MN}F^{MN} +\frac{i}{2}  \psi^T\hat C\hat \Gamma^MD_M\psi\right]\ ,
\end{align}
where  $M,N=1,2,3,...,10$,  the metric is $\hat \eta_{MN} = {\rm diag}({\mathbb I }_{5\times 5},-{\mathbb I}_{5\times 5})$ and
$$D_M\psi = \partial_M\psi - i[A_M,\psi], 
\qquad
F_{MN} = \partial_M A_N - \partial_N A_M - i [A_M,A_N] \, .$$
 We use a real basis for the $Spin(5,5)$ Clifford algebra with  $\{\hat\Gamma^M,\hat\Gamma^N\} = 2\hat \eta^{MN}$,  $\hat C\hat\Gamma^M\hat C^{-1} = -(\hat\Gamma^M)^T$ and a Majorana-Weyl spinor $\psi$, satisfying $\hat \Gamma_{11}\psi=-\psi$. The theory in 5+5 dimensions is of course not unitary, and we have chosen the sign of the lagrangian
so that it reduces to the same theory arising from timelike reduction of the (2,0) theory. The fields are Lie-algebra-valued, and the gauge fields $A_M$ are hermitian.

The dimensional reduction of this lagrangian is straightforward. However, to compare with formulae of other papers and aid with the 11D interpretation, we
 choose a representation of the  $Spin(5,5)$ Clifford algebra given in terms of
the   matrices of the $Spin (1,10)$  Clifford algebra with real generators $\Gamma^M$,  metric $\eta_{MN} = {\rm diag}(-1,1,...,1)$  and charge conjugation matrix $C =\Gamma_0$, satisfying
$\{\Gamma^M,\Gamma^N\} = 2 \eta^{MN}$.
A suitable choice for $\hat\Gamma^M$ is
\begin{align}
\hat \Gamma^i = \Gamma^i\qquad \hat\Gamma^I = -\Gamma_{12345}\Gamma^I\qquad \hat C = \Gamma_0\ ,
\end{align}
where $i,j=1,2,3,4,5$ and $I,J = 6,7,8,9,10$. In this case one finds
$\hat \Gamma_{11} = \Gamma_{012345}$ so that $\psi$  satisfies $\Gamma_{012345}\psi= -\psi$.

Dimensionally reducing gives
 the  following  action in 5 Euclidean dimensions:
\begin{align}\label{Eaction}
S    =  \frac{1}{g^2} \tr\int d^5x \Bigl[  & \frac{1}{4 }F_{ij}F^{ij} - \frac{1}{2} D_i\phi^ID^i\phi^I - \frac{1}{4}[\phi^I,\phi^J]^2\nonumber
\\ & - \frac{i}{2}  \psi^T\Gamma_0\Gamma^iD_i\psi
-\frac{1}{2} \psi^T \Gamma^I [\phi^I,\psi] \Bigr] \ .
\end{align} 
This action is invariant under the supersymmetry transformations
\begin{align}
\delta \phi^I  &= i \bar{\epsilon} \Gamma^I \psi \\
\delta \psi &= \Gamma^\mu \Gamma^I \epsilon D_\mu \phi^I  + \frac{1}{2 } \Gamma^{ij}\Gamma^0 \epsilon F_{ij  } + \frac{i}{2}  \Gamma_0 \Gamma^{IJ}  [ \phi^I, \phi^J]  \\
\delta  {A}_i &= -i   {\epsilon}^T \Gamma_i\psi\ ,
\end{align}
where $\Gamma_{012345}\epsilon= \epsilon$.
Note that   $\phi^I$, $A_i$,  $\psi$ and $g^2$ have mass dimensions  $1, 1, 3/2$ and $-1$, respectively. 

The kinetic terms for the gauge fields and the scalars have opposite signs: while that for the gauge fields is positive definite, that for the scalars is negative definite.
Multiplying the action by $-1$ would swap which bosonic term is positive and which is negative, but neither $S$ nor $-S$ is positive.
This is   a consequence of supersymmetry: there is no  action in 5D Euclidean space with 16 real supersymmetries whose bosonic part is real and positive definite.

The bosonic part of the
action is real but not positive. For the path integral we perform an analytic continuation $\phi^I\to i\phi^I$
to give an action whose bosonic part is real and positive definite. This is the same Euclidean action as that used for the Wick-rotation of the 4+1 SYM theory, where one Wick rotates time and takes $A_0\to -i A_0$.
Thus, as remarked in the introduction, the same Euclidean path integral is used for both the 4+1 SYM and the 5+0 SYM, but with different continuations back to the real section.

The SYM theory  (\ref{Eaction}) has  conserved currents associated with the symmetries of the action $S$. In particular, translational invariance gives rise to
\begin{align}\label{Tis}
\Theta_{ij} =&   \frac{1}{g^2}\tr \left(D_i \phi^I  D_j \phi^{I } - \frac{1}{2} \delta_{ij} D_k \phi^I   D^k \phi^{I } - \frac{1}{4} \delta_{ij}   [\phi^I,\phi^J]^2 - F_{ik} F_j{}^k +\frac{1}{ 4 } \delta_{ij}F_{kl}F^{kl}\right.
\nonumber\\
 &- \left.\frac{i}{2} \bar{\psi}  \Gamma_i D_j \psi  + \frac{i}{2} \delta_{ij} \bar{\psi}  \Gamma^kD_k\psi  + \frac{1}{2} \delta_{ij} \bar{\psi}  \Gamma_0\Gamma^I   [\phi^I,  \psi ]  \right)\ ,
\end{align}
and invariance under supersymmetry gives rise to the supercurrent
 \begin{align}\label{Sis}
{\cal S}^i =& \frac{1}{g   }\tr \left( -\frac{1}{2 } F_{jk  } \Gamma^{jk}\Gamma_0 \Gamma^i \psi  - D_j \phi^I  \Gamma^j\Gamma^I \Gamma^i \psi  +\frac{i}{2}   [ \phi^I,  \phi^J]  \Gamma_0 \Gamma^{IJ} \Gamma^i\psi    \right) \, .
\end{align}
In addition to these  we can also identify a \lq topological' current
\begin{align}\label{Jis}
J_{ i} = \frac{1}{8g^2}\varepsilon_{ijklm} \tr (F^{jk}F^{lm})\ .
\end{align}
which is identically conserved as a result of the Bianchi identities. We can also identify a positive-definite scalar quantity
\begin{equation}\label{His}
H = \frac{1}{g^2}\tr \Bigl(  \frac{1}{4 }F_{ij}F^{ij} + \frac{1}{2} D_i\phi^ID^i\phi^I - \frac{1}{4}[\phi^I,\phi^J]^2\Bigr)\ ,
\end{equation}
that we will relate to the energy shortly.

Let us now compare with the timelike reduction of the (2,0) theory in 5+1 dimensions.  We do not have a formulation of the non-abelian (2,0) theory. Although it is not expected to have a conventional field theory   formulation,   it is expected to have a local energy-momentum tensor $T_{\mu\nu} $.
 However
the abelian (2,0) theory is known explicitly;  the bosonic fields consist of a closed self-dual three-form $H_{\mu\nu\lambda}$ and 5 scalars $\Phi^I$. 
The energy-momentum tensor for the abelian (2,0) theory  has the bosonic terms
\begin{align}\label{Ta}
  T^{(abelian)}_{\mu\nu} &= {2\pi}\left(\frac{1}{2\cdot 2!}H_{\mu\lambda\rho}H_\nu{}^{\lambda\rho}  +\partial_\mu \Phi^I \partial_\nu \Phi^I-\frac{1}{2}\eta_{\mu\nu} \partial_\lambda \Phi^I \partial^\lambda \Phi^I\right )\ .
\end{align}
The normalization factor of $2\pi$ was explained in \cite{Maldacena:1997de} and we also justify it in an appendix.
The equations of motion can be deduced from the conservation equation
\begin{equation}
\partial^\mu T^{(abelian)}_{\mu\nu}=0 \ .
\end{equation}

Let us now   compactify on a timelike circle of radius $R$.  If we write 
\begin{align}
H_{0ij} = \frac{1}{g^{2}}F_{ij}\qquad H_{ijk}= \frac{1}{2g^2}\epsilon_{0ijklm}F^{lm}\qquad \Phi^I = \frac{1}{g^2} \phi^I\ ,
\end{align} 
where 
\begin{equation}
g^2 =  {4\pi^2 R}\ ,
\end{equation}
then
we find
\begin{align}
\int dt \, T^{(abelian)}_{ij} &=  \Theta_{ij}^{(abelian)}\\
\int dt \, T^{(abelian)}_{0i} &=   J_{i}^{(abelian)}\\
\int dt \, T^{(abelian)}_{00} &=   H^{(abelian)}\ . 
\end{align} 
Here $ \Theta_{ij}^{(abelian)}$, $J_{i}^{(abelian)}$ and $H^{(abelian)}$ are the abelian limits of $ \Theta_{ij} $, $J_{i} $ and $H $ given above.
While $J_i^{(abelian)}$ is  the  topological conserved current,  $\Theta_{ij}^{(abelian)}$ is the stress-tensor arising from the action
\begin{align}\label{Mactionb}
S   =  \frac{1}{g^2}\int d^5x \Bigl[   \frac{1}{4}F_{ij}F^{ij} - \frac{1}{2} \partial_i\phi^I \partial ^i\phi^I  \Bigr] \ .
\end{align}
Thus the time-like compactification of the abelian (2,0) theory gives the 5D abelian version of the action (\ref{Mactionb}). 

For the non-abelian (2,0) theory, the timelike reduction must give a 5D theory with this abelian limit. 
 Gauge invariance and supersymmetry then determines this 5D non-abelian theory to be the 5D SYM
 with action  (\ref{Eaction}) and stress-tensor (\ref{Tis}).
 Here we have considered a straightforward dimensional reduction with no KK modes. More generally, we will identify (\ref{Tis}), (\ref{Jis}) and (\ref{His}) as the Fourier zero-mode components of the full six dimensional energy momentum tensor $T_{\mu\nu}$.

\section{ Instantons and Kaluza-Klein Modes }

We begin by recalling how an extra space dimension arises in 4+1 dimensional SYM to give the (2,0) theory in 5+1 dimensions  \cite{Rozali:1997cb}, and then seek an analogous understanding of an extra time dimension in  our Euclidean 5D SYM.

The 4+1 SYM theory has BPS solitons given by the product of an  (anti-)self-dual Yang-Mills instanton in
4 dimensional Euclidean space  with a time dimension.
Integrating the conserved topological current $  J=\frac{1}{2g^2}*tr(F \wedge F)$ over a $4$-space at fixed time
gives the charge
\begin{align}
\label{topochg}
K&\equiv \int  d^4 x   J_0 = {4 \pi ^2 n\over  g^2}\ .
\end{align}
Assuming suitable fall-off in the transverse $\R^4 $, the  integer  $n$ is the instanton number and the charge is independent of time, as a consequence of the conservation of $J$. The simplest such 
instantonic solitons are half-supersymmetric and have mass given by
\begin{align}
M&=\vert K \vert\ .
\end{align}
As $g\to \infty$, these solitons all become massless and the spectrum becomes continuous.

A theory in 5+1 dimensions compactified on a spacelike circle in the $x^5$ direction has a quantized momentum in that direction
\begin{align}
P_5&={n\over R}\ ,
\end{align}
where $R$ is the radius of the circle and $n$ is an integer.
A massless field in 5+1 dimensions will then give a Kaluza-Klein tower of BPS states for all integers $n$ with mass
\begin{align}
M&=\vert P_5 \vert\ .
\end{align}
As $R\to \infty$, these all become massless and the spectrum becomes continuous. It was proposed in  \cite{Rozali:1997cb}  that the solitons in 4+1 dimensions arising from instantonic solitons
should be interpreted as Kaluza-Klein modes for a theory compactified on a circle of radius
\begin{align}
R&= { g^2 \over 4 \pi ^2
}\ ,
\end{align}
so that the strong coupling limit $g \to \infty$ is interpreted as a decompactification of a spatial dimension, $R\to \infty$. The BPS state with instanton number $n$ is interpreted as the $n$'th Fourier mode  on the $x^5$ circle.

There are two theories with 16 supersymmetries in 5+1 dimensions and no higher spin fields that are candidates for the theory arising in this strong coupling limit: the interacting (2,0) theory and the SYM theory with (1,1) supersymmetry.
One way of seeing that it is the (2,0) theory arising in this way is that the worldvolume theory for a stack of D4-branes should,
at strong string coupling, become  the
worldvolume theory for a stack of M5-branes wrapped on the M-theory circle.
Another argument arises from the supermultiplet structure. The theory in 5+1 dimensions should have 16 supersymmetries, but could have (1,1) supersymmetry or (2,0) supersymmetry. These two supersymmetry algebras have $SO(4)$ and $SO(5)$ R-symmetries respectively. In particular,
Kaluza Klein states for (1,1)  supersymmetry and Kaluza Klein states for (2,0)  supersymmetry fit into different  massive BPS multiplets of $N=4$ supersymmetry in 4+1 dimensions 
 \cite{Hull:2000cf}.
The solitons for 4+1 SYM fit into the same multiplets as the  Kaluza-Klein modes for (2,0) supersymmetry, so that the theory in 5+1 dimensions must have  (2,0) supersymmetry \cite{Hull:2000cf}.

For
the SYM theory in 4+1 dimensions it is straightforward to rule out the possibility that the extra dimension opening up at strong coupling is timelike. The timelike reduction of a superymmetric theory in 4+2 dimensions would have to give the standard SYM theory in 4+1 dimensions with $SO(5)$ R-symmetry. However, in 4+2 dimensions, there are no Majorana-Weyl fermions or real self-dual tensors, so the only multiplet with 16 supersymmetries and no higher spins is the SYM theory in 4+2 dimensions, which has R-symmetry $SO(3,1)$. Dimensional reduction on a timelike direction gives the wrong SYM theory -- it gives  the non-unitary SYM theory in 4+1 dimensions with R-symmetry $SO(4,1)$ that arose in \cite{Hull:1998vg,Hull:1999mt} (and which can also be constructed by compactifying from the SYM theory in 5+5 dimensions on 4 time dimensions and one space dimension).

 The current $J_\mu$ of the 4+1 SYM theory ($\mu=0,1,2,3,4$)
comes from the components $T_{\mu 5}$ of the energy-momentum tensor in 5+1 dimensions.
The charge $K$ then can be written as
\begin{align}\label{Kist}
K&=   \int  d^5 x  \, T_{05}\ ,
\end{align}
(with an integral over the 5 spatial coordinates, for fixed $x^0$)
and this supports the interpretation of $K$ as a component of the momentum, $P_5$.

We now seek a similar argument for the Euclidean SYM theory, essentially interchanging $x^0$ and $x^5$ in the above.
The Euclidean SYM has supersymmetric solutions 
that are non-trivial in 4 dimensional space but extended along an extra spatial dimension.
If the coordinate in this extra space dimension is $x^5$, then the solution consists of an  (anti-)self-dual gauge field $A_i$ in the 4D space with coordinates $x^1,x^2,x^3,x^4$ that is independent of $x^5$, with $\phi^I=0$.
Note that this solution is translationally invariant along $x^5$ and preserves $1/2$ of the supersymmetries, with supersymmetry parameter satisfying $\Gamma_{1234}\epsilon=\pm\epsilon$.
A localised instanton in $\R^4$ then gives rise to a 1 dimensional extended object in
$\R^5$.
Following \cite{Hull:1998vg}, we shall refer to a $p$ dimensional supersymmetric extended object in Euclidean space as an $Ep$-brane, so that this is an $E1$-brane.

  %
%

A theory in 5+1 dimensions compactified on a time-like circle in the $x^0$ direction is expected to have a quantized momentum in the timelike direction
\begin{align}
E &={n\over R}\ ,
\end{align}
where $R$ is the radius of the circle and $n$ is an integer. The spectrum of timelike Kaluza-Klein modes is then identical to the spectrum of K-charges if we relate the radius to the coupling constant by $R= { g^2 /4 \pi ^2}$.
We then propose that the Euclidean SYM theory (with a suitable UV completion) is  the timelike compactification of the (2,0) theory of radius
\begin{align}\label{Ris}
R&= { g^2 \over 4 \pi ^2 }\ .
\end{align}
The continuous spectrum of $K$-charges arising in the strong coupling limit $g \to \infty $ is then interpreted as the decompactification of a timelike circle.

Supersymmetry again dictates the signature of the extra dimension. The strong coupling limit cannot  involve a spatial dimension opening up to give a theory in 6 Euclidean dimensions, as there  is no theory in 6 Euclidean dimensions with 16 supersymmetries
that gives the required SYM theory on dimensional reduction.
In 6 Euclidean dimensions,
 there are no Majorana-Weyl fermions or real self-dual tensors, so the only multiplet with 16 supersymmetries and no higher spins is the SYM theory, which has R-symmetry $SO(3,1)$.
Dimensional reduction of this gives the wrong SYM theory -- it gives  the SYM theory in 5 Euclidean dimensions with R-symmetry $SO(4,1)$.

A theory with 16 supersymmetries and a timelike dimension opening up could give either a (1,1) theory or a (2,0) theory in 5+1 dimensions. The supermultiplet structure of the BPS states carrying the K-charge matches with that of the timelike Kaluza-Klein modes for a (2,0) theory but not for a (1,1) theory, so that the strong coupling limit must have (2,0) supersymmetry.
This can also be seen from an M-theory perspective, as we will discuss in section 8.

However, there are a number of issues raised by this discussion concerning the quantity $E$.
First, the instanton number and hence the K-charge can have either sign, so the identification with an energy $E $ would seem to require that $E$ can have either sign.
Moreover, this energy seems to know about the choice of $x^5$ direction whereas the usual energy in 5+1 dimensions is $SO(5)$ invariant and independent of any choice of spatial direction.
We have seen that
 the current $J_i$
comes from the components $T_{i0}$ of the energy-momentum tensor in 5+1 dimensions and that the charge $K$ then can be written as (\ref{Kist}).
  %
%
In the next section we address these questions   about our suggestion that $K$ be interpreted as a momentum around a timelike direction, {\it i.e.} as an energy in 5+1 dimensions.
The key point, as we shall see, is to regard $x^5$ as a Euclidean time and to analyse the theory in terms of a canonical approach adapted to $x^5$  instead of the usual one centred on dependence on time $x^0$. This was to be expected as we are trying to interchange the roles of $x^0$ and $x^5$ in the usual picture.

\section{Euclidean Time, Charges and Canonical Formulation }

Time plays a central role in quantum theory, so there are interesting issues about what it means to quantize a Euclidean field theory in $\R^{d}$. 
A standard approach is to choose one of the coordinates, $\tau$ say,  and treat this similarly to the way one treats time in the usual case.
Given a choice of $\tau$, the $\tau$ derivatives of fields can be eliminated in favour of conjugate momenta and a canonical formalism can be set up.  Poisson brackets can be defined in the usual way for fields at equal $\tau$, and
a Hamiltonian introduced to govern $\tau$ dependence. Symmetries lead to conserved charges, {\it i.e.} ones that are independent of $\tau$.
In a canonical quantization, the Poisson brackets become commutation relations for quantum operators at equal $\tau$.

It is natural to ask whether
 path integral quantization can avoid needing to choose a time. However, standard definitions of the path integral involve a time-slicing of paths in Minkowski space, or a slicing with respect to a Euclidean time for a
Euclidean path integral. It is this slicing that gives rise to the canonical commutation relations at fixed time or
 Euclidean time. It is interesting to ask whether there can be an alternative formulation of path integrals that doesn't involve such a slicing.
Formally, the path integral can be thought of as  generating Euclidean space correlation functions, which can be calculated 
in  perturbation theory in terms of Euclidean space Feynman
diagrams. The quantum theory might be thought of as being defined by these correlation functions, which will be covariant and seem to not depend on any choice of Euclidean time.
We will return to a discussion of this and other approaches to the quantization in Euclidean space in section 9.

The approach based on choosing a Euclidean time is natural if the Euclidean theory arises from Wick rotation from Minkowski space, with $\tau $ the analytic continuation of real time. 
In standard canonical quantization in $d+1$ dimensional Minkowski space, only the  $SO(d)$ spatial rotation symmetry is manifest, and  the $SO(d,1)$ Lorentz symmetry has to be established; this is rather non-trivial in the canonical approach.
For canonical quantization in  $d+1$ dimensional Euclidean space using a Euclidean time, the expected $SO(d+1)$ rotational symmetry is not manifest. There is a manifest  $SO(d)$ spatial rotation symmetry, and the issue of $SO(d+1)$ symmetry then has to be addressed; it may arise in a way analogous to the way Lorentz symmetry arises in standard canonical quantization.

In Minkowski space, a natural boundary condition is to demand that fields fall off sufficiently fast in a spacelike surface at fixed time.
In a Euclidean theory, the analogue would be to demand that fields fall off sufficiently fast in the transverse space  at fixed $\tau$. We saw in the last section that E1-branes in the $x^5$ direction satisfy such boundary conditions, with the fields falling off in the transverse space at fixed $x^5$. Thus for such configurations, one can choose the coordinate along the E1-branes ($x^5$ for the case of the last section) as the Euclidean time $\tau$.
A choice of $\tau $ direction would be appropriate for discussing configurations falling off in the directions transverse to that direction, which would include E1-branes pointing in that direction, along with any other extended states that extend along that direction, such as those we will encounter in sections 5 and 6.
This leads to a quantization for the theory with the boundary condition that fields fall off in the transverse space at fixed $\tau$.
Other configurations falling off in other transverse spaces, such as
E1-branes pointing in other directions, would need to be treated separately using other choices of $\tau $ direction.

We now focus on the  5D Euclidean Yang-Mills theory and choose one direction with coordinate $\tau$ that we will treat as the Euclidean time, together with four orthogonal coordinates $x^a$, $a=1,2,3,4$, so that $x^i=(\tau, x^a)$.
We restrict to configurations that fall off sufficiently rapidly in the transverse $\R^4$ of fixed $\tau$, and we require this for any fixed value of $\tau$.
 Given a conserved current $j_i$  (with $\partial _i j^i=0$)
 we define
\be
\hat q = \int d^4x\,  j^\tau\ ,
\ee
where the integral is over a spatial 4-surface of fixed $\tau$.
 With our boundary conditions, namely that the fields fall off sufficiently quickly in the transverse directions,
  $\hat q$ is conserved in the sense that
\be
\partial_\tau \hat q =0 \ .
\ee
The energy-momentum tensor $\Theta_{ij}$ then leads to charges
%
%
%
\begin{align}
\hat \Pi_\tau &=  \int d^4x\, \Theta_{\tau}{}^{ \tau} \\
\hat \Pi_a &= \int d^4 x \, \Theta_{a}{}^{ \tau}\ .
\end{align}
These are the conserved charges corresponding to the translation invariance in $\R ^5$, with $\hat \Pi_\tau$ the hamiltonian generating $\tau$ evolution. Introducing Poisson brackets of the canonical formulation based on  $\tau$,
these charges generate translations in $\R^5$ through the Poisson brackets:
\begin{align}
[\hat \Pi_i , \psi]  = \partial _i \psi\ ,
\end{align}
for any $\psi$.
The topological current $J$ leads to the E1-brane charge
\begin{align}
\label{kis}
K&=   \int  d^4 x\,  J^\tau\ ,
\end{align}
as before,
while the
supercurrents (\ref{Sis}) give the supercharges
\begin{align}
\hat Q = \int d^4 x \ {\cal S}^\tau\ .
\end{align}

It is straightforward to calculate the superalgebra generated by these supercharges using the $\tau$ Poisson brackets.
It can also be calculated using the method of
\cite{Hull:1983ap}
and  noting that
\begin{align}
[\bar\epsilon ^\alpha  \hat Q_\alpha , \hat Q_\beta ] =&  \int d^4x \ ( \delta_\epsilon {\cal S}^\tau )_{\beta}\ ,
\end{align}
It is  easier to calculate the supersymmetry variation of the super current $\delta_\epsilon {\cal S}_\tau$ than to calculate the Poisson brackets.
The superalgebra
is of the form
 \begin{align}
\{ \hat Q_\alpha , \hat Q_\beta \}
=& 2 ( \Gamma^i  C^{-1} )_{\alpha \beta} \hat \Pi_i - 2 \delta_{\alpha \beta}    K + \dots \, .
\end{align}
where various tensorial brane charges have been omitted.
It was already suggested  in the last section that the topological charge $  K$ be interpreted as a momentum in an extra time dimension. If we write $    \hat \Pi_0=K$, then the algebra can be written in a way that is suggestive of 5+1 dimensions
as
 \begin{align}
\label{hatsalg}
\{ \hat Q_\alpha , \hat Q_\beta \}
=&
2 ( \Gamma^\mu C^{-1} )_{\alpha \beta} \hat \Pi_\mu+\dots \, .
\end{align}
In fact the whole superalgebra can be organised into the same form as the superalgebra of the (2,0) theory in 5+1 dimensions,
with
\begin{align}
\{ \hat Q_\alpha , \hat Q_\beta \} =
& 2 ( \Gamma^\mu C^{-1} )_{\alpha \beta} \hat \Pi_\mu + ( \Gamma^\mu \Gamma^I C^{-1} )_{\alpha \beta} \hat {\cal Z}_\mu^{I} + ( \Gamma^{\mu \nu \lambda} \Gamma^{IJ} C^{-1} )_{\alpha \beta} \hat {\cal Z}_{\mu \nu \lambda}^{IJ} \, .
\end{align}
Here the right hand side has been expanded in terms of all possible symmetric $16\times 16$ matrices. One finds that (setting the fermions to zero for simplicity and using the equations of motion)
 \begin{align}\label{firstHQ}
  \hat \Pi_a &= \frac{1}{g^2}\tr \int d^4x D_a\phi^I D_\tau \phi^I-F_{ab}F_\tau{}^b\\
 \hat \Pi_\tau &= \frac{1}{g^2}\tr \int d^4x \frac{1}{4}F_{ab}F^{ab} - \frac{1}{2}F_{\tau a}F_\tau{}^a -\frac{1}{4}[\phi^I,\phi^J][\phi^I,\phi^J]\nonumber
 \\ 
 &\qquad\qquad\qquad + \frac{1}{2} D_\tau \phi^I D_\tau \phi^I - \frac{1}{2} D_a\phi^I D^a\phi^I
 \label{hatppis}
 \\
\hat \Pi_0&= \frac{1}{8g^2} \tr \int \varepsilon _{\tau bcde}  F^{bc}F^{de}   \\
  \hat {\cal Z}_0^{I} &= \frac{2}{g^2}\tr \int d^4x D^a( F_{\tau a} \phi^I)    \label{zcharge}
  \\
  \hat {\cal Z}_a^I &= -\frac{1}{g^2}\varepsilon_{\tau abcd}\tr \int d^4x \ D^b(F^{ cd }  \phi^{I } )   \\
    \hat {\cal Z}_\tau^I &= 0 \\
\hat {\cal Z}_{0 a \tau}^{IJ} &= -\frac{i}{3!\cdot 3!g^2}\varepsilon^{IJKLM}\tr\int d^4x \  D_a([\phi^K,\phi^L]\phi^M)  \\
\hat {\cal Z}_{0 a b}^{IJ} &= \frac{i}{4!g^2}\varepsilon_{\tau abcd}\tr\int d^4x \  F^{cd}[\phi^I,\phi^J] \\
\hat {\cal Z}_{ab\tau}^{IJ} &=\frac{1}{3g^2}\tr\int d^5x \   \frac{i}{4}F_{ab} [\phi^I, \phi^J]  - D_{[a}\phi^ID_{b]}\phi^J \, . \label{lastHQ}
 \end{align}
We see that the superalgebra indeed involves all the charges of a 5+1 dimensional theory.

We now turn to the charges of the (2,0) theory in 5+1 dimensions.
For configurations that satisfy the standard boundary conditions that the fields fall off asymptotically in a spatial $\R ^5 $ of constant $t=x^0$, the momenta have the standard form
\begin{align}
P_\mu = \int d^5 x \ T_{\mu}{}^{ 0}\  .
\end{align}
However, to make contact with the charges considered above in 5 Euclidean dimensions, it is useful to consider instead boundary conditions in which fields fall off asymptotically in a space of constant $\tau$, where $\tau $ is one of the spatial coordinates, {\it e.g.} $\tau = x^5$.
This will be appropriate only for a subsector of the theory, including  strings extended along the $\tau$ direction.
We will mostly be interested in the case of periodic time $x^0$, so that the space of constant $\tau$ is the product of Euclidean  $\R ^4$ with a timelike circle with period $2\pi R$. In that case, we will require fields to fall off asymptotically in the transverse
 $\R ^4$. Then we can define alternative charges as integrals over this $S^1\times \R ^4$
of constant $\tau$:
\begin{align}\label{P20}
\hat P_\mu &= \int d^4 x dt  \ T_{\mu}{}^{ \tau}\nonumber\\
&= \int d^4 x \, ( T_{\mu}{}^{ \tau})_0\ ,
\end{align}
where $ (T_{\mu\nu})_0$
is the zero Fourier mode of $T_{\mu \nu}$:
\begin{align}
(T_{\mu}{}^{\nu})_n =    \int dt \, e^{-in t/R} T_{\mu}{}^{\nu}\ .
\end{align}

For these boundary conditions, one could introduce a canonical formalism based on the coordinate $\tau$, instead of the conventional one based on $t$.
We conjecture that at strong coupling $g$, the 5D Euclidean SYM theory becomes the (2,0) theory quantized using $\tau$ as Euclidean time and compactified on a timelike circle of radius $g^2/4\pi^2$, with the charges $\hat \Pi_i$ of the 5D theory giving the spatial components  $\hat P_i$ of (\ref{P20}),
so that 
\begin{equation}
\hat P_i=   \hat \Pi_i\qquad \hat P_0= \hat \Pi_0=  K\ , 
\end{equation}
and also $\hat Z^I_\mu = \hat {\cal Z}^I_\mu$, $\hat Z^{IJ}_{\mu\nu\lambda} = \hat {\cal Z}^{IJ}_{\mu\nu\lambda}$.
This is what follows from interchanging time $t$ and the Euclidean time $\tau$ in the standard relation between 4+1 SYM and the (2,0) theory. 
Furthermore we see from (\ref{hatppis}) that
\begin{equation}
\hat \Pi_i = \int d^4x \Theta_{i}{}^{ 5}\qquad
K= \int d^4x J^5\ .
\end{equation}
Since the choice of direction of $\tau = x^5$ was arbitrary we conclude more generally that $\Theta_{ij}$, $J_i$ and $H$ can be identified with the Fourier zero-modes of $T_{ij}$, $T_{0i}$ and $T_{00}$ respectively;
\begin{align}
\Theta_{ij} &= \int dt \, T_{ij} \\
J_i & = \int dt \, T_{0i}\\
H & = \int dt \,T_{00}\ ,
\end{align}
 in agreement with the abelian results in section 2. The supersymmetry algebra then requires us to also identify
\begin{equation}\label{TTis}
  \hat {\cal Z}^I_\mu =  \int dt \hat Z^I_\mu\qquad    \hat {\cal Z}^{IJ}_{\mu\nu\lambda} =   \int dt \hat Z^{IJ}_{\mu\nu\lambda}\ .
\end{equation}

Lastly we turn to the relation between the charges $\hat P_\mu$   and the conventional charges $P_\mu$ in the 5+1 dimensional theory. 
We first consider  string-like configurations that are independent of both $t$ and $\tau= x^5$, such as the strings
that we will consider in sections 5 and 6.
The momenta $P_\mu$ are interpreted as charges per unit length,  and the
$\hat P_\mu$ are interpreted as charges per unit time, both
given  as integrals over a transverse $\R^4$ at some constant values $t=t_0$, $\tau= \tau_0$:
\begin{align}
{P_\mu}  = \int d^4 x \ T_{\mu}{}^{ 0}\  , \qquad \qquad   \hat P_\mu    &=   
 \int d^4 x \ T_{\mu}{}^{ \tau}\  ,
\end{align}
and hence
\begin{align}\label{match}
\hat P^0 &=    P_\tau 
 \ .
\end{align}

Alternatively, without assuming translational invariance along $t$ and $\tau $,  we can consider a boosted sector where the fields only depend on $x^a$, $a=1,2,3,4$ and $x^5+vt$ (or a null sector where $v=\pm 1$). Then  by a simple change of variables
one finds a similar relation to (\ref{match})
\begin{align}\label{boostedmatch}
 {\hat P^0}   =
 \int d^4 x dt  \ T^{0 \tau} =
  \frac{1}{  |v|}
\int d^4 x d\tau \ T_{\tau}{}^{ 0}=  
 \frac{1}{  |v|}
P_\tau \ .
\end{align}

We see that for such sectors of the theory, the time component of the momentum $ \hat P_0$
resulting from the topological charge $K$ in 5-D
is identified with the $\tau $ component of the conventional momentum $P_\mu$, and this of course need not be positive.
This resolves the issues raised at the end of the last section.
Comparing the 5+1 theory on a timelike circle with the 5D Euclidean SYM theory at strong coupling, the 5D charges $(\hat \Pi _ i, K)$ lift to the charges
$  
\hat P_\mu$ in 5+1 dimensions, and these can be related to the conventional charges
$  P_\mu$ for certain sub-sectors of the theory. We also expect that similar relations to (\ref{boostedmatch}) arise in the case of other charges obtained from scalar integrands.

\section{ Matching of BPS States}

In this section and the next,  we compare the spectrum of supersymmetric  states of the 5D Euclidean SYM with that of  the (2,0) theory in 5+1 dimensions. The BPS states of the 5D SYM can be extrapolated to strong coupling where they can be identified with the BPS states of the (2,0) theory.
Specifically, we will consider   classical BPS solutions of 5D SYM and seek to match these with
BPS states of the (2,0) theory. 
 In the decompactification limit $R\to \infty$, these BPS states should fit into representations of $SO(5,1)$, and we will seek evidence for such  $SO(5,1)$ covariance in the 5D BPS spectrum.

The usual 6D $(2,0)$ superalgebra takes the form
\begin{align}
\{  Q_\alpha ,  Q_\beta \}
=&  2 ( \Gamma^\mu C^{-1} )_{\alpha \beta}   P_\mu + ( \Gamma^\mu \Gamma^I C^{-1} )_{\alpha \beta}   Z_\mu^{I} + ( \Gamma^{\mu \nu \lambda} \Gamma^{IJ} C^{-1} )_{\alpha \beta}   Z_{\mu \nu \lambda}^{IJ} \, .
\end{align}
A similar algebra is satisfied by the hatted charges arising in the canonical formalism based on $\tau = x^5$, as discussed in the previous section:
\begin{align}
\{  \hat Q_\alpha ,  \hat Q_\beta \}
=&   2 ( \Gamma^\mu C^{-1} )_{\alpha \beta}   \hat P_\mu + ( \Gamma^\mu \Gamma^I C^{-1} )_{\alpha \beta}   \hat Z_\mu^{I} + ( \Gamma^{\mu \nu \lambda} \Gamma^{IJ} C^{-1} )_{\alpha \beta}  \hat Z_{\mu \nu \lambda}^{IJ} \, .
\end{align}
In the conformal phase the central charges vanish and the 1/2 BPS states of the (2,0) theory in 5+1 dimensions consist of massless states with null momentum $P_\mu P^\mu=0$. In the Coulomb branch, where conformal invariance is spontaneously broken, there are static self-dual strings carrying the charge $Z_i^I$ \cite{Howe:1997ue}. 
There are further  1/4 BPS states, some of which were  considered in  \cite{Lambert:2010iw}.

The supersymmetric states of SYM in 5 Euclidean dimensions
are all extended along (at least) one direction -- there are no point-like BPS states. As before, we choose  a coordinate $\tau$ along such a direction and coordinates $x^a$ for the the transverse $\R^4$ orthogonal to it.
For configurations falling off sufficiently fast in the transverse $\R^4$, we can define momenta $\hat P_i$ (\ref{P20}) and a topological charge $K$ (\ref{kis}) together with   electric charges
\begin{align}
\label{qis}
q^I& =   \oint tr (\langle \phi^I \rangle *F)\ ,
\end{align}
where the integral is over the $S^3$ at infinity in the transverse $\R^4$.
In this section and the next,
 we will   be interested  in expectation values $\langle \phi^6 \rangle$   of $\phi^I$ that break the gauge group to an abelian subgroup.
The charge $q^I$  arises in the super-algebra (\ref{hatsalg}) as the charge $ \hat Z_0^{I} $  in  (\ref{zcharge}).

For configurations extended along   2 dimensions with fields falling off asymptotically on a transverse $\R^3$, one can define a magnetic charge
\begin{align}
p^I& =   \oint tr (\langle \phi^I \rangle F)\ ,
\end{align}
integrated over the $S^2$ at infinity in the transverse $\R^3$. This   does not appear
in the superalgebra (\ref{hatsalg}).
\footnote{ It does occurs as a charge in the superalgebra of the 5D Euclidean SYM theory with R-symmetry $SO(4,1)$.}


\subsection{Massless States in 5+1}

A massless  6D state with null momentum with, say, $P^0 = |P^5|$ and all other charges vanishing is 1/2 BPS. 
Assuming that 
  the state only depends on $x^1,..,x^4$ and $x^5 \pm t$, corresponding to particles moving at the speed of light on the trajectory $x^5=\pm x^0$,
  we can also consider the canonical charges based on using $\tau = x^5$ as the \lq time'.
   We find that the solution carries
the 6D charges $\hat P_5$ and $\hat P_0$ 
with $\hat P_5= \pm \hat P_0$,  and  it is  also 1/2 BPS from the point of view of the hatted superalgebra.
As expected from the discussion in section 4, equation (\ref{boostedmatch}), we find
\begin{align}
  P_5&= \int d^4 x dx^5 \, T_5{}^{0} = \int d^4x dt T^{05} = \hat P^0  ,
\end{align}

If $x^5$ is compactified on a  circle of radius $R$ and $x^0$ is non-compact, then momentum is quantized with $P_5=n/R$ and the trajectory $x^5=\pm x^0$ spirals around the cylinder parameterised by $x^0$ and the periodic coordinate $x^5$.
On the other hand if $x^0$ is compactified and $x^5$ is not, we expect $\hat P_0$ to be quantized with $\hat P_0=n/R$ (by the standard arguments with $x^0$ and $x^5$ interchanged). This is to be associated with the curve
$x^5=\pm x^0$ which spirals around the cylinder parameterised by $x^5\in \R$ and the periodic coordinate $x^0$. In particular since time $x^0$ is only defined modulo $2\pi R$, $x^5$ is multivalued and the particle appears in many places at once.  Such acausal trajectories are solutions of the equations of motion and we shall suppose that they are permitted here.

The corresponding supersymmetric states in 5D have already been discussed in sections 3 and 4.
In 4+1 dimensions, corresponding to a spacelike compactification, there are instantonic solitons given by the product of
an  (anti-)self-dual gauge field $A_i$ in the $x^1,x^2,x^3,x^4$ plane with time to get a time-like worldline.
The mass $P^0$ is given by $|K|$ where $K= {4 \pi ^2 n/ g^2}$ is the topological charge (\ref{topochg}) and the momentum by $P_5= K$.
 
In Euclidean 5D SYM, corresponding to a timelike compactification, 
we expect a solution carrying the charges $K$ and $\hat \Pi_5$.
There is indeed such a
BPS solution    consisting of an  (anti-)self-dual gauge field $A_a$ in the $x^1,x^2,x^3,x^4$-plane that is independent of $x^5$, with $\phi^I=0$.
This solution is translationally invariant along $x^5$ and preserves $1/2$ of the supersymmetries where $\Gamma_{1234}\epsilon=\pm\epsilon$.
This is the instantonic E1-brane, and in terms of the 5D charges of sections 3,4 with $\tau = x^5$, it carries the momentum $\hat \Pi_0 =K$, $\hat \Pi_5 =|K|$   where $K$ is the instanton charge  (\ref{kis}).
This is then associated with the BPS states in 5+1 dimensions with periodic $x^0$, with
\begin{equation}
{\hat P}_0 =K\qquad {\hat P}_5 = \hat \Pi_5 \ .
\end{equation}

A  localized state that is traveling along $x^5$ at the speed of light in six-dimensions with worldline $x^0=\pm x^5$
then  corresponds to a 0-brane in 4+1 dimensional SYM with worldline in the $x^0$ direction or to an E1-brane in 5+0 dimensional SYM in the $x^5$ direction. In both cases, the momentum in the compact dimension arises from the instanton
charge. 

\subsection{1/2 BPS Static String Ground States}

We now turn to the 1/2 BPS states in the Coulomb phase in 5+1 dimensions that are the self-dual  static string ground states  carrying charge $Z^I_i$. There is a manifest $SO(5)$ rotational symmetry so we can fix the string to lie along $x^5$. There is also a manifest $SO(5)$ R-symmetry that acts on the $I$-indices and therefore we can fix $I=6$. These solutions are static so that  only $P^0$ and $Z^6_5$ are non-vanishing and given as charge densities per unit length along $x^5$ by    integrals over a transverse $\R^4$ with fixed $x^5$.  From the supersymmetry algebra one sees that $P^0 = \frac{1}{2}|Z^6_5|$ with the preserved supersymmetries satisfying $\Gamma^{056}\epsilon=\pm\epsilon$.

If compactified on a spacelike circle in the $x^5$ direction,  the string   winds the circle to give an electrically charged  0-brane in 4+1 dimensions, or if  compactified on a spacelike circle in another direction, {\it e.g.}
the $x^4$ direction, then
 it gives a magnetically charged string in 4+1 dimensions.
For a timelike circle, the 1+1 dimensional world-sheet of the string must wrap the time dimension to leave an E1-brane in 5 Euclidean dimensions.

As these states are extended along the $x^5$ direction, we consider
the canonical charges based on using $\tau = x^5$ as the \lq time'.
In terms of hatted charges this means that $\hat P_0=P_5=0$ with $\hat P_5$  non-vanishing.  
Therefore in 5D Euclidean SYM
we look for solutions with  $\hat \Pi_0=0$ but   non-vanishing $\hat \Pi_5$. In addition, from the superalgebra we also see that preserving the supersymmetries  $\Gamma^{056}\epsilon=\pm\epsilon$ now requires that $\hat Z^6_5=0$ and  $\hat \Pi_5= \pm\frac{1}{2}\hat Z^6_0 \ne 0$.   
  We then seek co-dimension four  solutions n 5D Euclidean SYM with $F_{ab}=0$ but  $F_{a5}$ non-zero, $a,b=1,2,3,4$. They are   `electric' E1-branes with
\begin{align}
\label{eleo}
F_{a5}  = \pm D_a\phi^6\ .
\end{align}
It is easy to see that they are 1/2 BPS with
$\Gamma_{05}\Gamma^6\epsilon=\mp\epsilon$.
The solution for a single string located at the origin is simply
\begin{align}\label{sdssol}
\phi^6 = \langle \phi^6 \rangle - \frac{g^2Q_E}{4\pi^2 r^2}\ ,\qquad A_5 = \pm \phi^6\ ,\qquad A_a=0\ ,
\end{align}
where 
$$r^2= \sum _{a=1}^4 (x^a)^2.$$

The hatted-momentum per unit length along $x^5$ satisfies
\begin{align} 
\hat P_5 = \hat \Pi_5 = \pm\frac{1}{2}\hat {\cal Z}^6_0\ .
\end{align} 
However the explicit expressions are divergent due to the singularity at $r=0$. 
The integrals over the transverse space $\R^4$ diverge, and when rewritten as surface integrals, 
 in addition to the boundary at infinity there is also an inner boundary given by  some small sphere of radius $r=\epsilon$.
 The surface integral at infinity is finite, but the integral over the inner boundary is divergent.
Just as in the case of fundamental strings ending on D-branes,  this divergence arises because these solutions represent semi-infinite M2-branes that come in from infinity in the $\phi^6$ direction and end on the M5-branes leading to a string-like defect along $x^5$.  The infinite energy is then due to the infinite area of the M2-branes \cite{Callan:1997kz} (see the appendix).
 It would be interesting to compare these solutions to the work of \cite{Chu:2013hja,Chu:2012rk}.


\subsection{Magnetically Charged Solutions}

In the last subsection we considered electrically charged states. There will also be magnetically charged states obtained by lifting 3D  magnetic monopole solutions. We consider these here, and find that surprisingly they are not supersymmetric (although more general supersymmetric monopole solutions will play a role in what follows).
These then lift further to magnetically charged 2-brane classical solitons in 5+1 dimensions. These are non-superymmetric, but may be protected by a fake supersymmetry in the sense of \cite{Skenderis:2006fb}.

There are magnetically charged  solutions of 5D Euclidean SYM that have a non-trivial profile along $x^m$, $m=1,2,3$.
For magnetic charge $p^6$,
these are given by monopole solutions in $x^m$  with
\begin{align}
\label{momo}
F_{mn}& =  \varepsilon_{mnp}  D^p F\qquad \phi^6 = \varphi\ .
\end{align}
Here we have introduced a scalar   function $\varphi$ which
 satisfies $D^mD_m \varphi=0$ as a result of the Bianchi identity.
Solutions to these non-linear equations are well-known as BPS monopoles and can be obtained using the Nahm construction.  One finds a moduli space of solutions  
that are parameterised by the asymptotic value of $\varphi$ at infinity:
\begin{align}
\varphi = \langle \varphi \rangle +\ldots\ ,
\end{align}
and the magnetic charge $p^6=2\pi \tr (\langle \varphi \rangle Q_M)$ where
\begin{align}
Q_M =   \frac{1}{2\pi}\oint  F\ ,
\end{align}
is the magnetic flux integrated over the spatial 2-sphere at infinity.

These solutions  are independent of $x^4,x^5$ and so constitute an E2-brane extended over the
$x^4,x^5$ directions.
It is straightforward to check that these are not supersymmetric solutions of the SYM with R-symmetry $SO(5)$ that we are studying here, but that they are 1/2 BPS E2 brane  solutions of the 5D Euclidean
SYM with R-symmetry $SO(4,1)$.
These monopole E2 branes of the
SYM with R-symmetry $SO(5)$
might be thought of as having fake supersymmetry.

These results about the supersymmetry can also be understood as follows.
The SYM theory in 3+1 dimensions with R-symmetry $SO(6)$ has 1/2 BPS monopole solutions.
Reducing on time gives 1/2 BPS solutions of the 3D Euclidean SYM theory with R-symmetry $SO(6,1)$ of the form (\ref{momo}). These can then be lifted to
  1/2 BPS solutions of the   Euclidean SYM theory in $K$ dimensions ($K\le 9$) with R-symmetry $SO(9-K,1)$. These are independent of $x^4,..., x^K$ and so constitute $E(K-3)$ branes extended over $x^4,..., x^K$. On the other hand, the 5D Euclidean SYM theory with R-symmetry $SO(5)$ we have been focusing on here reduces to a 3D Euclidean SYM theory with R-symmetry $SO(5,2)$, and the monopole solution of the form (\ref{momo}) does not preserve any supersymmetries of this theory.
 Its $E2$ lift to 5D then preserves no supersymmetries. This is just as well, as there seem to be no BPS states in the (2,0) theory in 5+1 dimensions that could correspond to a BPS E2 brane in 5-D.

\subsection{1/4 BPS Excited String  States}

Let us now consider  string states in 5+1 dimensions that carry string charge $Z^6_5$ in the $x^5$ direction
and momentum $P_5$ parallel to the string.
This cannot be 1/2 BPS because a string in its ground state cannot carry momentum along its length.
 In particular boosting a ground state of a string along its length has no effect because it preserves such Lorentz symmetries. Thus a string carrying momentum along its length must be in an excited state and hence have at most 1/4 of the supersymmetry.

Since the string remains fixed in space we expect that the five dimensional solution will be extended along $x^5$ but localized in the remaining four dimensions.
It should carry instanton charge $K$ corresponding to a non-zero $\hat P_0= P_5$ in addition to the electric charge 
$\hat {\cal Z}^6_0$ we obtained above.
The required solutions are \lq dyonic instantons'  \cite{Lambert:1999ua} consisting of an anti-self-dual gauge field in the transverse $\R^4$ with coordinates $x^a$ where $a=1,...,4$ as well as a non-vanishing electric-type gauge field given by
\begin{align}
 F_{a5} = \pm D_a \phi^6 \ .
\end{align}
The  preserved supersymmetries  satisfy
\begin{align}
0&= \left(  \Gamma^a\Gamma^6D_i\phi^6  \pm  \Gamma^{a5}\Gamma^0F_{a5}+\frac{1}{2}\Gamma^{ab}\Gamma^0F_{ab} \right)\epsilon\ .
\end{align}
This is solved by  requiring $\Gamma_{1234}\epsilon=\epsilon$ and $\Gamma^6\Gamma_{05}\epsilon= \mp\epsilon$. Note that since $\Gamma_{012345}\epsilon=\epsilon$ the first condition is equivalent to $\Gamma_{05}\epsilon=\epsilon$ and the second to $\Gamma^6\epsilon=\mp\epsilon$.

The gauge field $A_a$ is determined by the ADHM construction and, for each   instanton number,   there is a moduli space of solutions. In addition
the equations of motion then require that $D^2\phi^6=0$. Given an instanton gauge field and asymptotic value $\langle \phi^6 \rangle $ for $\phi^6$, there is a unique solution  for $\phi^6$ of $D^2\phi^6=0$ which will have  the asymptotic form
\be
\phi^6 = \langle \phi^6 \rangle \mp \frac{g^2 Q_E}{4\pi^2 r^2}  +\ldots \ .
\ee

As above, the charges can be formally defined as integrals over the ($x^1,x^2,x^3,x^4$)-plane giving momenta {\it etc.} per unit length in the $x^5$ direction.
\begin{align}
Q_E  = \frac{1}{ g^2}\oint F_{r5}
\end{align}
is the electric flux integrated over the 3-sphere at infinity. Note that $Q_E$ is determined by the instanton moduli and $\langle \phi^6 \rangle$  and hence is not a free parameter \cite{Lambert:1999ua}. 

Taking $\tau = x^5$ and evaluating the hatted charges  (\ref{firstHQ})-(\ref{lastHQ}) we find
\begin{align}
 \hat P_0=
K
  =  \frac{ n }{R} \qquad
    \hat P_5 = 
  {  \hat \Pi_5}  = \left| K \right| -\frac{1}{2}\left| {\hat {\cal Z}}^6_0 \right|\qquad 
   {\hat {\cal Z}^6_0}  = -2\tr(\langle \phi^6 \rangle Q_E)\ .
\end{align} 
We note that the absolute values arise because we must have 
\begin{align}
0\le \tr \int (D\phi^6)^2 =\tr \oint \phi^6 D_r \phi^6 
 = \pm \tr(\langle \phi^6\rangle Q_E)\ ,
\end{align}
and also
\begin{align}
  \tr \int F_{ab}F^{ab} =\frac{{\rm sgn} (n)}{2}\tr \oint \epsilon^{abcd}F_{ab}F_{cd} =  16\pi^2 |n|\  .
\end{align}

\section{Boosted String States}

We have argued that 5D SYM  should be identified with the $(2,0)$ theory compactified on a timelike circle. In this section, we seek further evidence for this conjecture by looking for
some signature of Lorentz symmetry in 5+1 dimensions. In particular we would like to look for solutions of the 5D theory that can be interpreted as boosted versions of the solutions we considered in the previous section. In a 6D theory  with Lorentz symmetry one expects to find states that fill out representations of $SO(5,1)$.

At first one might think that this is not possible in theories with compact time, as one would  expect Lorentz invariance only to be recovered in the limit $g^2 \to \infty$. In particular
compact time 
 restricts   $\hat P^0$   to be discrete: $ \hat P^0 = n/R$, $n\in {\mathbb Z}$ while boosts depend on a continuous parameter $v$ and do not preserve the discreteness of  $\hat P^0$. However if the 6D theory has a conformal symmetry, even if it is spontaneously broken, then one can compensate for the continuous shift in $ \hat P^0$ by also including a conformal transformation.

In particular, consider a state in 5+1 dimensional Minkowski space with $ \hat P_4=0$ and then boost it along $x^4$. This leads to a state with
\begin{align}
 \hat P'_0 = \gamma \hat P_0\ ,\qquad \hat P'_4 = \gamma  v \hat P_0\ ,  \qquad   \hat P_\mu'= \hat P_\mu \quad \mu\ne 0,4 \ .
\end{align}
Now perform a conformal rescaling $\hat P_\mu''=\gamma^{-1}\hat P'_\mu$ to obtain
\begin{align}\label{boostedC}
\hat  P''_0 =   \hat P_0\ ,\qquad \hat P''_4 =    v \hat P_0\ ,  \qquad \hat  P_\mu''=\gamma^{-1} \hat P_\mu  \quad \mu\ne 0,4 \ .
\end{align}
Then for   states with $ \hat P_4=0$, we can perform a continuous boost along $x^4$ combined with a conformal transformation and preserve $ \hat P^0$. In this way we get a spectrum of states with the same  $ \hat P^0$ and  depending on a continuous parameter $v$. This means that if we now compactly time, $t=x^0$,  on a circle of radius $R$, then for a given discrete momentum $ \hat P^0 = n/R$ we should expect a spectrum of states labelled by a continuous parameter $v$. In this section, we seek to find the corresponding solutions of 5D Euclidean SYM,  depending on a continuous parameter $v$, providing further evidence that the SYM theory is really a 5+1 dimensional theory.
 We will refer to these states as boosted although it is important to keep in mind that they have also been conformally rescaled. 
This action on the momenta also applies to any 6D vector such as the central charges $\hat Z^I_\mu$, so we would obtain
\begin{align}\label{boostedZ}
\hat  Z''_0 =   \hat Z_0\ ,\qquad \hat Z''_4 =    v \hat Z_0\ ,  \qquad \hat  Z_\mu''=\gamma^{-1} \hat Z_\mu  \quad \mu\ne 0,4 \ .
\end{align}

Note that the in the case of spatial compactifications this does not work for the conventional (unhatted) charges. In particular if $  P_5$ is discrete then we can clearly consider Lorentz transformations that leave $   P_5$ invariant: {\it e.g.} a boost along $x^4$ gives
\begin{align}\label{boost}
P'_0 = \gamma  (P_0+vP_4)\ ,\qquad P_4 = \gamma (P_4+ v P_0)\ ,  \qquad    P_\mu'=  P_\mu \quad \mu\ne 0,4 \ .
\end{align}
Including boosts along $x^1,x^2$ and $x^3$ generates the $SO(1,4)$ of the theory dimensionally reduced on $x^5$. If we instead consider a boost along $x^5$ and conformal rescaling by $\lambda$ we find
\begin{align}
  P''_0 = \lambda\gamma (P_0+vP_5) \ ,\qquad P''_5 = \lambda\gamma(P_5+ vP_0)\ ,  \qquad P_\mu''=\lambda P_\mu \quad \mu\ne 0,5 \ .
\end{align}
In this case, the  construction  analogous to that above would be to keep $P_5''=P_5$ fixed by taking $P_0=0$ and $\lambda = \gamma^{-1}$. But in a unitary theory only the ground states   satisfy $P_0=0$ and they are simply boost invariant.

\subsection{ Boosted 1/2 BPS String Ground States}

Let us now consider  a charged string in 5+1 dimensions in the $x^5$ direction, which carries momentum perpendicular to $x^5$, say along $x^4$. It corresponds to a string in its ground state that has been boosted in a transverse direction, so that it is
a 1/2 BPS  state. It carries charges $\hat P_0, \hat P_4$ and $\hat Z_5^6$, and there is a natural action of $SO(1,1)$ acting as boosts in the 4 direction. Combining the boost with a conformal rescaling will preserve $ \hat P^0$, so that if $ \hat P^0 = n/R$ this spectrum of states depending continuously on $v$ will still be present if $x^0$ is compactified on a circle of radius $R$.

We now seek corresponding supersymmetric states of the 5D SYM theory, and expect to see a spectrum of such states depending continuously on $v$. The solution will clearly be extended along $x^5$.
The 5D solution corresponding to the unboosted string is the electric E1 brane of section 6.2.
For the boosted string, the momentum $\hat P_4$ is expected  arise from an instantonic E1 brane in the 4 direction, corresponding to instantons   in the  $x^1,x^2,x^3, x^5$ plane, independent of $x^5$.
 Following the example of the pure momentum mode above, the solutions expected to be extended along   $x^4$ as well as $x^5$. Thus we should look for a 1/2 BPS solution that has a non-trivial profile along $x^m$, $m=1,2,3$.  These are given by monopole solutions in $x^m$. We consider the following ansatz:
\begin{align}
F_{mn} = \varepsilon_{mnp} D^p\varphi \qquad  A_5= \frac{1}{v}\varphi  \qquad  \phi^6= \pm\frac{1}{v\gamma  }\varphi\ ,
\end{align}
with the only non-vanishing fields being $A_m,A_5, \phi^6$. Here $\varphi$ is the usual scalar field that appears in the BPS monopole system and satisfies $D^mD_m \varphi=0$ as a result of the Bianchi identity, and all fields depend only on
 $x^m$. The parameters $v,\gamma$ are to be determined but their form has been chosen for future convenience. Note that if $v=1$, the fields $A_m, A_5$ satisfy $F_{mn} = \varepsilon_{mnp} F_{p5}$ so that the solution would be  an   instanton in the $x^1,x^2,x^3, x^5$-plane that is independent of $x^5$.
Note also that $F_{m5} =  {\gamma } D_m \phi^6$, which is a rescaled version of (\ref{eleo}).

Supersymmetry is preserved if
\begin{align}
0&= \left(\frac{1}{2}\Gamma^{mn}\Gamma^0\varepsilon_{mnp}D^p\varphi + \frac{1}{v}\Gamma^{p5}\Gamma^0D_p\varphi\varphi \pm \frac{1}{v \gamma}\Gamma^p\Gamma^6 D_p\varphi\varphi\right)\epsilon\ .
\end{align}
Now $\frac{1}{2}\Gamma^{mn }\varepsilon_{mnp}\Gamma^0\epsilon = -\Gamma_p\Gamma^{45}\epsilon$ and so we require
\begin{align}
\left(\frac{1}{v}\Gamma^{04}\mp\frac{1}{v\gamma  }\Gamma^{456}\right)\epsilon =\epsilon\ .
\end{align}
This is a 1/2 BPS projector if $\gamma^2=1/(1-v^2)$ and in what follows we take $\gamma=1/\sqrt{1-v^2}>0$.
Note that for this solution we require $v\ne 0 $; below we shall match this solution with a string in 5+1 dimensions boosted to a velocity $v$.

To continue we observe that
\be
\langle  \varphi\rangle  =  \pm{v\gamma  } \langle \phi^6 \rangle  \ ,
\ee
so that $\phi^6\to \langle \phi^6 \rangle $ as $r\to \infty$ (with $r^2= x^mx^m$).
With our conventions   there is a topological quantization condition \cite{Goddard:1976qe}
\begin{align}
e^{i\oint F } =  e^{2\pi i   Q_M} = 1\ .
\end{align}
Such solutions  carry the magnetic charge
\begin{align}
p^6&=2\pi  \tr(\langle \phi^6 \rangle Q_M)\ .
\end{align}

Let us evaluate the hatted-charges per unit length for these solutions. The 6D solution is a function of $x^1,x^2,x^3$ and $x^4+vt$. Since it is periodic in $t$ it must also be periodic in $x^4$ with period $2\pi v R$ and therefore we have 
\begin{equation}\label{PtoPi}
\hat P_i = \int d^3x dx^4-Dt T_{i5} = \int d^3x dx^4 \Theta_{i5}   = 2\pi R v \hat \Pi_i 
\end{equation}
where $\hat \Pi_i$ is evaluated as an integral over $x^1,x^2,x^3$. Similarly we have
\begin{equation}
\hat P_0 = 2\pi Rv \hat \Pi_0\qquad \hat Z^I_\mu = 2\pi Rv\hat {\cal Z}^I_\mu \ .
\end{equation}
The resulting charges are  
\begin{align}
\label{chargess}
{\hat P}_5&= \mp\frac{ 1  } {\gamma } \tr(\langle \phi^6 \rangle Q_M)\qquad {\hat Z}^6_0  =-2 \tr(\langle \phi^6 \rangle Q_M) \qquad {\hat Z}^6_4  =   -v \tr(\langle \phi^6 \rangle Q_M)\ .
\end{align}
These are transform just as in (\ref{boostedC}) and satisfy
\begin{align}
 ({\hat P}_5)^2 =-\frac{1}{4}({\hat Z}^6_0)^2 +  \frac{1}{4}({\hat Z}^6_4)^2   \ ,
\end{align}
although neither the left-hand-side nor right-hand-side are independent of $v$ but rather scales with a factor of $\gamma^{-2}$ as a result of the conformal transformation discussed above.

\subsection{Boosted 1/4 BPS  Excited String  States}

Finally we wish to consider the case of an excited string state that also carries momentum
 perpendicular to $x^5$, say along $x^4$. Thus
we should also look for boosted and rescaled versions of the dyonic instantons solutions which had ${\hat \Pi_0}$, ${\hat \Pi_5}$ and ${\hat {\cal Z}^6_0}$ non-vanishing. According to our discussion above we should find    boosted and rescaled versions of these along $x^4$. Thus these solutions will be extended along $x^4$ and $x^5$, 1/4 BPS and have non-vanishing 
$K$, ${\hat \Pi_4}$, ${\hat \Pi_5}$ as well as ${\hat {\cal Z}^6_0}$ and ${\hat {\cal Z}^6_4}$ but ${\hat {\cal Z}^6_5}=0$. To find them we can consider 1/4 BPS monopoles that are functions of $x^m$, $m=1,2,3$ (for example see \cite{Bergman:1997yw,Lee:1998nv}) with $A_4,A_5$ and $\phi^6$ non-vanishing and non-commuting:
\begin{align}
0&= \left(\frac{1}{2}\varepsilon_{mnp}F^{np}\Gamma^{m45} +  D_mA_4\Gamma^{m40}+  D_mA_5\Gamma^{m50} +  D_m\phi^6\Gamma^{m6}\right)\epsilon\nonumber\\
& -i \left( [A_4,A_5]\Gamma^{450} +[A_4,\phi^6] \Gamma^{46} +[A_5,\phi^6] \Gamma^{56}\right)\epsilon\ .
\end{align}
 To solve this we write
\begin{align}
\phi^6 =  vA_4\pm\frac{1}{\gamma } A_5\ ,
\end{align}
to find
\begin{align}
0&= \Gamma^{i45}\left(\frac{1}{2}\varepsilon_{mnp}F^{np} +  D_mA_4(\Gamma^{50}-v\Gamma^{456})-   D_mA_5(\Gamma^{40}\pm\frac{1}{\gamma}\Gamma^{456})\right)\epsilon\nonumber\\
& -i [A_4,A_5]\left(\Gamma^{450}  \pm\frac{1}{\gamma} \Gamma^{46} -v \Gamma^{56}\right)\epsilon\ .
\end{align}
We must therefore impose the supersymmetry projector
\begin{align}
\left(\pm\frac{1}{\gamma} \Gamma^{560} +v\Gamma^{460}\right)\epsilon=\epsilon\ ,
\end{align}
which requires we take $\gamma =1/\sqrt{1-v^2}$ as before. This leaves us with
\begin{align}
0&= \Gamma^{i45}\left(\frac{1}{2}\varepsilon_{mnp}F^{np} \mp\frac{1}{\gamma} D_mA_4\Gamma^6 + v D_m A_5 \Gamma^6\right)\epsilon\ .
\end{align}
Thus we find a second projector $\Gamma^6\epsilon=\epsilon$ and BPS equation
\begin{align}
F_{mn} = \frac{1}{2}\varepsilon_{mnp}D^p\varphi\ ,
\end{align}
where
\begin{align}
\varphi=\pm\frac{1}{\gamma} A_4-v A_5\ .
\end{align}
The Bianchi identity for $F_{mn}$ implies
\begin{align}
 D^mD_m\varphi=0\ .
\end{align}
We also obtain another equation comes from the equation of motion for $\phi^6$:
\begin{align}\label{Xeq}
 D^mD_m\phi^6=[\varphi,[\varphi,\phi^6]]\ .
\end{align}
A standard argument shows that, given $\varphi$ and $A_m$ satisfying the BPS monopole equation, solutions  $\phi^6$ of this equation are uniquely determined by the asymptotic value $\langle \phi^6\rangle$.

We only want to look a solutions with  ${\hat {\cal Z}_5^6}=0$. Some algebra shows that this gives the condition
\begin{eqnarray}
0&=&\tr \oint  D_rA_4 \phi^6\nonumber\\
&=& \tr\oint \phi^6\left(\pm\frac{1}{\gamma}D_r \varphi +v D_r\phi^6\right)  \ .
\end{eqnarray}
Next we note that from (\ref{Xeq}) we can deduce that
\begin{eqnarray}
0=\tr\int \varphi D_mD^m \phi^6 &=&\tr\oint \varphi D_r \phi^6 - \tr\int D_m\varphi D^m\phi^6 \nonumber\\
&=& \tr\oint \varphi D_r \phi^6 - \tr\oint  \phi^6D_r \varphi\ .
\end{eqnarray}
Therefore the condition $\hat {\cal Z}^6_5=0$ becomes
\begin{align}
\tr\oint \left(\pm\frac{1}{\gamma}\varphi+v\phi^6\right)D_r\phi^6=0\ .
\end{align}
To continue we write
\begin{align}
\langle \varphi\rangle &= \pm\frac{4\pi^2 \gamma}{v g^2}\varphi_0 \mp v\gamma \langle \phi^6 \rangle \ ,
\end{align}
where $\varphi_0$ is to be determined. With this convention  the condition $\hat {\cal Z}^6_5=0$ becomes
\begin{align}
\tr\oint \varphi_0D_r\phi^6=0\ .
\end{align}
Since $\phi^6$ is determined by $\langle \varphi\rangle$, $\langle \phi^6\rangle$ and $Q_M$ this give a constraint on $\varphi_0$ in terms of $\langle \phi^6\rangle$ and $Q_M$. In particular it asserts that $\varphi_0$ must be orthogonal, in the Lie-algebra,  to the scalar flux
\begin{align}
Q_X = \frac{1}{4\pi}\oint D_r \phi^6 \ .
\end{align}
Using  $\tau= x^5$ and the same relation (\ref{PtoPi}) we find that this solution carries the hatted charges:
\begin{align}
\label{charges5}
 {\hat P}_0  &=  \frac{   1 }{ R} \tr(  \varphi_0  Q_M) \\
 {\hat P}_4  &=  \frac{  v}{R}\tr(  \varphi_0  Q_M)  \\
 {\hat P}_5  &= \pm \frac{1 }{\gamma R} \tr( \varphi_0   Q_M) \pm \frac{1}{\gamma  }\tr(\langle \phi^6 \rangle Q_M) \\
 {\hat Z}^6_0     &= 2\tr(\langle \phi^6 \rangle Q_M)  \\
 {\hat Z}^6_4    &=  2 v \tr(\langle \phi^6 \rangle Q_M) \ .
\end{align}
Again these charges are precisely of the form in (\ref{boostedC}) and reduce the the boosted 1/2 BPS states when $\varphi_0=0$. In addition requiring a discrete spectrum for $\hat P_0 = n/R$ imposes the quantization condition   $\tr(\varphi_0Q_M)\in {\mathbb Z}$.

\section{ 5D SYM and the $(2,0)$ Superalgebra}

There is an alternative way to
view the 5D SYM theory and its relation to a theory in 5+1 dimensions
  using the construction given in  \cite{lambert:2010wm} of a non-abelian system of equations that furnish a representation of the 6D $(2,0)$ superalgebra. We will see that this gives the same identification of the 6D $T_{\mu\nu}$ in terms of currents of 5D SYM that we derived in section 2.
  
 \subsection{Derivation of 5D SYM from the $(2,0)$ Superalgebra}
  
 Let us briefly review the  construction of  \cite{lambert:2010wm}. It contains 5 scalars $\Phi^I_a$, a sixteen-component fermion $\psi_a$ which satisfies $\Gamma_{012345}\psi_a=-\psi_a$,  a gauge field $ A_\mu{}^a{}_b$, a vector $C^\mu $ and a self-dual three-form $H_{\mu\nu\lambda\ a}$:
\begin{align}
H_{\mu\nu\lambda \; a} &= \frac{1}{3!}\epsilon_{\mu\nu\lambda\omega\sigma\rho}H^{\omega\sigma\rho}{}_a\; .
\end{align}
 Here the index $a$ refers to the fact that the fields take values in a Lie-algebra with structure constants $f^{ab}{}_c$.
Note that the original paper \cite{lambert:2010wm} used a three-algebra and took $C^\mu$ to have a 3-algebra index. However that is not necessary and for simplicity we have dropped it here. In particular we make the replacement $C^\mu_a f^{abc}{}_d\to C^\mu f^{bc}{}_d$
in the formulae in  \cite{lambert:2010wm}, where $f^{abc}{}_d$ are the structure constants of a Lie 3-algebra.

The supersymmetry transformations are:
\begin{align}
\delta \Phi^I_a &= i \bar{\epsilon} \Gamma^I \psi_a \\
\delta \psi_a &= \Gamma^\mu \Gamma^I \epsilon D_\mu \Phi^I_a + \frac{1}{3!} \frac{1}{2} \Gamma_{\mu \nu \lambda} \epsilon H^{\mu \nu \lambda}_a - \frac{1}{2} \Gamma_\lambda \Gamma^{IJ} \epsilon C^\lambda  \Phi^I_c \Phi^J_d f^{cd }{}_a \\
\delta H_{\mu \nu \lambda \; a} &= 3i \bar{\epsilon } \Gamma_{[\mu \nu} D_{\lambda]} \psi_a + i \bar{\epsilon} \Gamma^I \Gamma_{\mu \nu \lambda \kappa} C^\kappa  \Phi^I_c \psi_d f^{cd }{}_a \\
\delta  {A}_\mu{}^b{}_a &= i \bar{\epsilon} \Gamma_{\mu \lambda} C^\lambda \psi_d f^{db}{}_a \\
\delta C^\mu  &= 0 \, ,
\end{align}
where $\bar \epsilon = \epsilon^TC$ with $C=\Gamma_0$. These transformations close on-shell. In particular the equations of motion are \cite{lambert:2010wm}:
\begin{align}
0 &= \Gamma^\mu D_\mu\psi_a+\Phi^I_c C^\nu  \Gamma_\nu\Gamma^I\psi_d f^{cd }{}_a \\
0 &=D^2 \Phi_a^I -\frac{i}{2}\bar\psi_c C^\nu  \Gamma_\nu\Gamma^I \psi_d f^{cd }{}_a + C^\nu  C_{\nu  } \Phi^J_c \Phi^J_e \Phi^I_f f^{ef }{}_{d}f^{cd }{}_a \\
0 &= D_{[\mu}H_{\nu\lambda\rho]\;a}+\frac{1}{4}\epsilon_{\mu\nu\lambda\rho\sigma\omega}C^\sigma  \Phi^I_c D^\omega \Phi^I_d f^{cd }{}_a + \frac{i}{8}\epsilon_{\mu\nu\lambda\rho\sigma\omega}C^\sigma  \bar\psi_c \Gamma^\omega \psi_d f^{cd }{}_a \\
0&= F_{\mu\nu}{}^b{}_a - C^\lambda  H_{\mu\nu\lambda\; d}f^{ db}{}_a \\
\label{Ccon}
0 &= D_\mu C^\nu\\
\label{pconstraint}0 &= C^\rho  D_\rho \Phi^I_d  = C^\rho  D_\rho \psi_d  =C^\rho  D_\rho H_{\mu\nu\lambda\; a}  \ .
\end{align}
In all these equations $  F_{\mu\nu}{}^b{}_a$ is the field strength of the gauge connection $ A_\mu{}^b{}_a$ which appears in the covariant derivative $D_\mu$ which acts as, for example,  $D_\mu \Phi^I_a = \partial_\mu \Phi^I_a -   A_\mu{}^b{}_a\Phi^I_b$ and hence $F_{\mu\nu}{}^b{}_a = \partial_\mu A_\nu{}^b{}_a  - \partial_\nu A_\mu{}^b{}_a +A_\mu{}^b{}_cA_\nu{}^c{}_a - A_\nu{}^b{}_cA_\mu{}^c{}_a$.\footnote{This definition of $F$ differs by a sign from that used in \cite{lambert:2010wm,lambert:2011gb}}
These equations admit conserved currents \cite{lambert:2011gb} (including the $2\pi$ normalization of \cite{Maldacena:1997de}):
\begin{align}
\nonumber
\frac{1}{2\pi} \tilde \Theta _{\mu \nu} =& D_\mu \Phi^I_a D_\nu \Phi^{Ia} - \frac{1}{2} \eta_{\mu \nu} D_\lambda \Phi^I_a D^\lambda \Phi^{Ia} \\
\nonumber
	&+ \frac{1}{4} \eta_{\mu \nu} C^\lambda  \Phi^I_a \Phi^J_c C_{\lambda } \Phi^I_f \Phi^J_e f^{cd a} f^{ef }{}_d + \frac{1}{4} H_{\mu \lambda \rho \; a} H_{\nu}{}^{\lambda \rho \; a} \\
	&- \frac{i}{2} \bar{\psi}_a \Gamma_\mu D_\nu \psi^a + \frac{i}{2} \eta_{\mu \nu} \bar{\psi}_a \Gamma^\lambda D_\lambda \psi^a + \frac{i}{2} \eta_{\mu \nu} \bar{\psi}_a C^\lambda  \Phi^I_c \Gamma_\lambda \Gamma^I \psi_d f^{acd} \, \\
	\nonumber \\
\frac{1}{{2\pi}} \tilde {\cal S}^\mu =& \frac{1}{2} \frac{1}{3!} H_{\nu \lambda \rho \; a} \Gamma^{\nu \lambda \rho} \Gamma^\mu \psi^a - D_\nu \Phi^I_a \Gamma^\nu \Gamma^I \Gamma^\mu \psi^a - \frac{1}{2} C^\nu  \Phi^I_c \Phi^J_d \Gamma_\nu \Gamma^{IJ} \Gamma^\mu \psi^a f^{cd}{}_a \, .
\end{align}
Here and in what follows we use a tilde to denote quantities for this (2,0) field theory, to distinguish them from the corresponding quantities for the (2,0) M5-brane world-volume  theory. The two theories are of course closely related, as discussed in  \cite{lambert:2010wm,lambert:2011gb}.

The field equation (\ref{Ccon}) imposes that $C^\mu$ is covariantly constant, so that the theory has vacua with $C^\mu$ a constant vector and all other fields vanishing.
The cases of spacelike and null expectation values for $C^\mu$ were analysed in \cite{lambert:2010wm} and \cite{lambert:2011gb} respectively.
For a constant $C^\mu$ pointing in the $x^5$ direction, the field equations imply that all fields are
independent of $x^5$ and the theory becomes the SYM theory in 4+1 dimensions.
In the null case, where we introduce light-cone coordinates $(x^1,x^2,x^3,x^4,x^+,x^-)$ and take $C^\mu$ along the $x^+$ direction, then the system reduces to one dimensional motion on the moduli space of  an(anti-) self-dual gauge field on $(x^1,...,x^4)$ with $x^-$ playing the role of time.

Here we wish to consider the case where $C$ is a timelike vector, 
\be
C^\mu = g^2\delta^\mu_0\ .
\ee
In this case the fields must be time-independent and the theory reduces to a theory in 5 Euclidean dimensions. Let us  introduce  matrix-valued 5D fields $\phi^I = g^2\Phi^I_aT^a$, $\psi = g^2\psi_aT^a$, where $[T^a,T^b] = if^{ab}{}_cT^c$. In particular one finds $A_{\mu}{}^b{}_a = -A_{\mu c}f^{cb}{}_a$ and $F_{\mu\nu}{}^b{}_a = -F_{\mu\nu c}f^{cb}{}_a$. The  equations of motion  are
\begin{align}
0 &= \Gamma^iD_i\psi -i \Gamma_0\Gamma^I[\phi^I,   \psi]  \\
0 &=D^iD_i X ^I +\frac{1}{2}  [\psi^T   ,\Gamma^I \psi]   +   [\phi^J, [\phi^J, \phi^I] ]  \\
0 &= D_{[i}H_{jkl] }-\frac{i}{4} \epsilon_{0ijklm}  [\phi^I, D^m\phi^I] + \frac{1}{8}\epsilon_{0ijklm}   [\psi^T ,\Gamma_0\Gamma^m \psi] \\
0&= F_{ij} + g^2 H_{ij0}\ ,
\end{align}
where it is understood that $F_{ij}$ now acts in the adjoint, \ie\ $D_\mu \phi^I = \partial_\mu \phi^I - i[A_\mu,\phi^I]$, $[D_\mu,D_\nu]\phi^I = -i [F_{\mu\nu},\phi^I]$.
 Contracting the third equation with $\frac{1}{3!}\varepsilon^{0ijkln}$ and using the fourth equation leads to
\begin{align}
-\frac{1}{g^2}D_iF^{in}  + i [\phi^I,D^n\phi^I] - \frac{1}{2}  [\psi^T,\Gamma_0\Gamma^n\psi]=0\ ,
\end{align}
as expected. In fact these are just the equations of motion that arise from the action (\ref{Eaction}).

Formally we can introduce the momenta and supercharges   given by
\begin{align}
\tilde \Pi_\mu =  \int d^5 x \  \tilde  \Theta_{\mu}{}^0\ ,\qquad\qquad \tilde Q =   \int d^5 x \ \tilde {\cal S}^0\ .
\end{align}
We can then define a formal superalgebra obtained by the supersymmetry variation of the supercharge
\begin{align}
[\epsilon ^\alpha     \tilde Q_\alpha , \tilde Q_\beta ] =&  \int d^5x \ ( \delta_\epsilon \tilde{\cal S}^0 )_{\beta}\ .
\end{align}
This  gives the superalgebra of
 \cite{lambert:2011gb}:
\begin{align}
\label{salgg}
\{ \tilde Q_\alpha , \tilde Q_\beta \}
%
=& 2 ( \Gamma^\mu C^{-1} )_{\alpha \beta} \tilde \Pi_\mu + ( \Gamma^\mu \Gamma^I C^{-1} )_{\alpha \beta} \tilde {\cal Z}_\mu^{I} + ( \Gamma^{\mu \nu \lambda} \Gamma^{IJ} C^{-1} )_{\alpha \beta} \tilde {\cal Z}_{\mu \nu \lambda}^{IJ} \, .
\end{align}
In particular, using the equations of motion, we find the following charges (setting the fermions to zero): 
\begin{align}
\label{firstQ} \tilde\Pi^0 &= \frac{2\pi  }{g^4}\tr\int d^5x \frac{1}{4}F_{ij}F^{ij} + \frac{1}{2} D_i\phi^ID^i\phi^I - \frac{1}{4}[\phi^I,\phi^J]^2\\
\tilde \Pi _i& = -\frac{2\pi}{8g^4}\tr\int d^5x \,  \varepsilon_{0ijklm}F^{jk}F^{lm}\ ,\label{2ndQ}
\end{align}
and the central charges are 
\begin{align}
\tilde  {\cal Z}_0^{I} &= 0 \\
\label{zii}
 \tilde {\cal Z}_i^I &= \frac{4\pi}{g^4}\tr \int d^5x \  D^j (F_{ij}\phi^{I })    \\
\tilde {\cal Z}_{0 i j}^{IJ} &= -\frac{2\pi}{3g^4}\tr\int d^5x \ \frac{i}{4}F_{ij} [\phi^I,\phi^J] + D_{[i} \phi^I D_{j]}\phi^J\\
\tilde {\cal Z}_{k l m}^{IJ} &= -\frac{  2\pi i}{4! g^4}\varepsilon_{0klmij}\tr\int d^5x \  F^{ij} [\phi^I, \phi^J]   \, .\label{lastQ}
   \end{align}
However, for a timelike   $C^\mu$, the fields are all independent of time on-shell,    so that it would be problematic to regard this as a Poisson bracket algebra in the usual dynamical sense.
Nevertheless the energy momentum tensor $\tilde \Theta_{\mu\nu}$ agrees with the expressions found above for 5D SYM, up to a factor of  $2\pi/g^2 = 1/2\pi R$:
\begin{align}\label{tildeThatT}
\tilde \Theta_{ij} = \frac1{2\pi R} \Theta_{ij}\qquad 
\tilde \Theta_{0i} =  \frac1{2\pi R} J_i\qquad
\tilde \Theta_{00} & =\frac1{2\pi R} H\ .
\end{align}

\subsection{$P_\mu$ and $Z^I_\mu$ for the  BPS states}

We would now like to investigate the physical interpretation of $\tilde \Pi_\mu$. The $\tilde \Pi_\mu$ are not associated with a canonical formulation but rather are simply evaluated as integrals over all space and time. 
Nevertheless, at the level of classical solutions, the results of the previous section suggest that  $\tilde \Pi_\mu$ are related to the momenta of the (2,0) theory, quantized using Lorentzian time using (\ref{ThetaTis}).
In particular, 
following the interpretation of $\Theta_{ij}$, $J_i$ and $H$ as Fourier zero-modes of $T_{\mu\nu}$, from (\ref{tildeThatT}) we   identify
    \begin{equation}\label{ThetaTis}
    \tilde \Theta_{\mu\nu} =\frac{1}{2\pi R} \int dt \, T_{\mu\nu}\ .
    \end{equation}
We  also assume a similar relation for the central charges
\begin{equation}\label{ZZis}
  \tilde {\cal Z}^I_\mu =\frac{1}{2\pi R}  \int dt Z^I_\mu\qquad  \tilde {\cal Z}^{IJ}_{\mu\nu\lambda} =\frac{1}{2\pi R}  \int dt Z^{IJ}_{\mu\nu\lambda}\ .
\end{equation}
Let us see what this leads to for the various solutions that we constructed in sections 5 and 6.

First we consider the null momentum mode arising from an (anti-)self-dual gauge field in the $(x^1,x^2,x^3,x^4)$-plane. If we assume the associated state in the $(2,0)$ theory depends only on $x^a$, $a=1,2,..,,4$ and $x^5\pm t$, we find the momenta  are given by
\begin{align}\label{Pone}
P_0&= \int d^4 x dx^5 \, T_{0}{}^{0} = \int d^4x dt T_{0}{}^{0}  = 2\pi R\int d^4 x \tilde \Theta_{0}{}^{0} \ ,
\\
 \label{Ptwo} P_5&= \int d^4 xdx^5 \,  T_{5}{}^0 = \int d^4 x dt\,  T_{5}{}^0 = 2\pi R\int d^4 x \tilde \Theta_{5}{}^0\ . 
\end{align}
This leads to 
\begin{align}\label{Ppi}
P^0 = \frac{ |n|}{R}\qquad P_5 =  -\frac{ n}{R}\ .\end{align}
in agreement with the discussion in section 5.1.

Next we look at the 1/2 BPS string ground states which consist of singular electric states extended along $x^5$. As before the charges per unit length are singular but satisfy
\begin{align}
{P^0} = \frac 1 2 \left\vert  { Z^6_5} \right \vert .
\end{align}
These are abelian solutions and hence one can use the known abelian (2,0) theory to compute their energy and charges per unit length.  
In the appendix we give a further analysis of the divergence to support the claim that these states correspond to semi-infinite M2-branes that intersect the M5-branes along a string-like defect.

Next consider the  1/4 BPS string states carrying $P_5$ momentum corresponding to smooth dyonic instantons. As with the null momentum mode we assume that the associated state in the $(2,0)$ theory depends on $x^a$, $a=1,2,..,,4$ and $x^5\pm t$. A similar argument to  that for the pure instanton above again leads to the relations (\ref{Pone}) and (\ref{Ptwo}) and hence we find  the charges 
\begin{align} 
  { P^0}  &= \left| { P_5}{}\right| +\frac{1}{2}\left|{Z^6_5}{}\right|\\
  {  P_5} &= -\frac{n }{R}\\ 
 {Z^6_5} &=   -2  \tr(\langle \phi^6 \rangle Q_E) 
\ .
\end{align} 

Let us now look at the boosted solutions.
Since these have been discussed in detail in the preceding section  from the point of view of the hatted charges we will be brief here and simply look at the 1/4 BPS states as the 1/2 BPS cases arise as special cases when $\tr (\varphi_0Q_M)=0$. The  corresponding 5+1 dimensional states are taken to be functions of $x^1,x^2,x^2$ and $x^4+vt$. For these solutions
the momenta per unit length can be evaluated as in (\ref{PtoPi}):
\begin{equation}
P_\mu  = \int d^3 x dx^4  T_{\mu}{}^{ 0} = |v|\int d^3x dt T_{\mu}{}^{ 0} = 2\pi R |v| \int d^3x \tilde \Theta_{\mu}{}^{ 0}
\end{equation}
where the integral is only evaluated  over $x^1,x^2,x^3$.  Similarly for the central charges. This results in the charges per unit length  
 \begin{align}
   {P^0}   &=    \gamma \left( \left|{   {P}_5}\right| +\frac{1}{2}\left|{{ Z}^6_5}\right|\right) \\
{P_4}  &=      v \gamma \left( \left|{  {P}_5}\right| +\frac{1}{2}\left|{ Z}^6_5\right|\right)    \\
{P_5}&=  -\frac{1   }{2\pi R^2 }  \tr(\varphi_0Q_M)  \\
{{Z}_5^6}&=\mp\frac{1}{\pi R }\left|\tr(\langle \phi^6 \rangle  Q_M)\right|  \ .
 \end{align}
The absolute value signs arise because  one can deduce the  bounds $\mp v\gamma\tr(\langle \phi^6 \rangle  Q_M)\ge 0$ and $\pm v\gamma\tr(\varphi_0  Q_M)\ge 0$  by evaluating the inequality $\tr\int (D\phi^6+\zeta DF)^2\ge 0$ for arbitrary $\zeta\in \R$.

 We now consider further the physical interpretation of these solutions.
 We have considered two quantizations of the (2,0) theory. One is the standard   quantization using the usual time  dimension and the other  is a quantization using a Euclidean time dimension.  These   are closely related as they are obtained from each other using a double Wick rotation. We can consider such a transformation of the theory in 6D Minkowski space but also when compactified on a Lorentzian torus, with both time and Euclidean time periodic with periods $2\pi R$ and $2\pi R'$ respectively. 
 Furthermore, any supersymmetric classical solution that is compatible with the boundary conditions appropriate for both quantizations should lead to a BPS state in both quantum theories. In particular,   string states  whose world-sheets     wrap both time and Euclidean time should be mapped to each other.
 Therefore it is natural to   expect that at least some classes of solutions in 5D Euclidean SYM, which as we have discussed can be interpreted as states in the Euclidean time quantization of the (2,0) theory, can also be interpreted as states in the standard  Lorentzian quantization of the (2,0) theory.

 The usual quantization   features  the $P_\mu$ charges whereas the Euclidean time quantization involves the $\hat P_\mu$ charges. 
 If we examine the $P_\mu$ charges of  the BPS states given above we see that they    give what one would expect from those of the  5+1 dimensional (2,0) theory,  compactified on a spacelike circle of radius $R=g^2/4\pi^2$. In particular $P_5$ is discrete and the boosted solutions are compatible with the expected $SO(1,4)$ Lorentz symmetry of a 5+1 dimensional theory compactified on a spacelike circle. Therefore  our results suggest that 5D SYM can also describe  some states and momenta of the (2,0) theory with non-compact time but compactified but on a spacelike circle (which in turn is described by 4+1 dimensional SYM).   
  
One might view this as a Lorentzian analogue of S-duality in the following way. First consider the familiar example  of the (2,0) theory  compactified on a Euclidean torus to 3+1 dimensional SYM, where the coupling constant and theta-angle give the modular parameter of the torus. Weak coupling corresponds to a degeneration where one torus direction decompactifies compared to the other. However, modular symmetry of the torus relates this to the opposite picture where the other radius is large, leading to  strongly coupled  3+1 dimensional SYM. Thus modular symmetry leads S-duality of 3+1 dimensional SYM, exchanging weak and strong coupling expansions of the same theory. One also expects S-duality to  hold for the (2,0) theory on a finite-sized torus, corresponding to  a  symmetry of 4+1 dimensional SYM on a circle of finite radius that relates Kaluza-Klein modes to solitons. Next consider the (2,0) theory on a Lorentzian torus with periods $2\pi R$ and $2\pi R'$ for time and Euclidean time respectively.  This gives a 4D Euclidean SYM with R-symmetry $SO(5,1)$ coupled to a tower of Kaluza-Klein modes. 
The modulus of the torus now lives in the coset space $SL(2,\R)/SO(1,1)$.
The  symmetry under large diffeomorphisms of the torus give rise to an $SL(2,\Z )$ S-duality symmetry of the 4D SYM, relating weak and strong coupling.
In the limit where one circle decompactifies (and the other does not) it is natural to quantize the theory using that direction as `time' regardless of whether it is Euclidean or Lorentzian. Depending on the signature of this `time' direction one either finds 5D SYM or 4+1-SYM. Thus in this sense a certain sector of states in these two theories are related by  Lorentzian $SL(2,\Z )$ transformations. 

In particular    we see that the double Wick rotation discussed above  takes  states of the Euclidean 5D SYM with coupling $g^2 = 4\pi^2 R$ compactified on a spacelike circle of radius $2\pi R'$ (corresponding to the Euclidean quantization of the (2,0) theory compactified on a Lorentzian torus)   to states of Lorentzian 5D SYM with coupling $g^2=4\pi^2 R$ compactified on a timelike circle of radius $2\pi R'$ (corresponding to the Lorentzian quantization of the (2,0) theory on a Lorentzian torus). The results of this section then provide evidence for this relation in the limit $R'\to\infty$.

\section{M-Theory, Branes and Time}

SYM in 4+1 dimensions (with higher derivative corrections) arises as the worldvolume theory for a stack of D4-branes in the IIA string, with Yang-Mills coupling given by
   \begin{align}
g^2 & = (2\pi )^2 {\alpha ' }^ {1/2} g_s\ ,
\end{align}
with $g_s$ the string coupling, so that going to strong-coupling in the SYM theory necessarily involves going to  strong coupling in the IIA string theory.
At strong coupling, the IIA string becomes 11D M-theory compactified on a spatial circle of radius
   \begin{align}
   \label{Mrad}
R & =  {\alpha ' }^ {1/2} g_s\ ,
\end{align}
and the D4-branes become M5-branes wrapped on the M-theory circle.
This implies that the strong coupling limit of the D4-brane worldvolume theory
becomes the M5-brane worldvolume theory at strong coupling, compactified on a circle or radius
\begin{align}   \label{YMrad}
R&= { g^2 \over 4 \pi ^2 \
}\ ,
\end{align}
which of course agrees with the relation (\ref{Ris}) obtained earlier.
The M5-brane theory has   higher derivative couplings, but has a decoupling limit giving the (2,0) superconformal field theory in 5+1 dimensions.

The conjecture that the strong coupling limit of the IIA string is 11-dimensional M-theory then implies that the D4-brane worldvolume theory must gain an extra dimension at strong coupling, and conversely the conjecture that the 4+1 SYM gains an extra dimension at strong coupling would imply that the IIA string theory too should gain an extra dimension.
In this section we discuss the corresponding situation for Euclidean theories and extra time dimensions.

Timelike dimensional reduction of 11-dimensional supergravity gives a supergravity in 10 Euclidean dimensions, the $IIA_E$ theory of  \cite{Hull:1998ym}. There are many problematic issues concerning quantum theories with periodic time, some of which are discussed in
 \cite{Hull:1998vg, Hull:1998ym} and references therein.
If we assume that it makes sense to put M-theory on a timelike circle, then for small radius
this formally defines a string theory in 10 Euclidean dimensions with coupling constant given in terms of the radius by (\ref{Mrad}). This is the $IIA_E$ string theory of  \cite{Hull:1998ym} with fundamental branes with 2 Euclidean dimensions (from wrapping M2 branes on time) and with a field theory limit given by the $IIA_E$ supergravity.
A stack of M5-branes, with 5+1 dimensional worldvolume, necessarily wrap the compact time dimension and give a stack of branes of the $IIA_E$ theory with 5 Euclidean dimensions, the $E5$ branes \cite{Hull:1998fh}.
The M5-brane worldvolume theory reduces to a  Euclidean 5D SYM theory on the $E5$ branes, and this is precisely the SYM theory considered here
\cite{Hull:1999mt}. The SYM coupling constant is again given in terms of the radius by (\ref{YMrad}).

It was conjectured in  \cite{Hull:1998ym} that the strong coupling limit of the $IIA_E$ theory is given by M-theory on a timelike circle, as suggested by the construction above. In particular, it has supersymmetric states corresponding to the Kaluza-Klein modes of M-theory on a timelike circle.
The strong string coupling limit implies a strong coupling limit of the $E5$ brane worldvolume SYM theory. As the $E5$ branes of the $IIA_E$ theory become M5-branes wrapped on the timelike circle at strong string coupling, this implies that the strong coupling limit of the Euclidean SYM theory on the $E5$ branes must be the (2,0) M5-brane worldvolume theory on a timelike circle. Conversely, the conjecture that the strong coupling limit of the 5D Euclidean SYM is (2,0) theory on a timelike circle provides strong evidence for the strong coupling limit of the $IIA_E$ string theory being M-theory on a timelike circle.

Much of the discussion of the Euclidean SYM in this paper also applies to the $IIA_E$ theory.
The $IIA_E$ theory is an analytic continuation of the usual IIA theory, and they are both formally  governed
by the same Euclidean path integral. The fact that the $IIA$ theory gains an extra dimension at strong coupling implies that the   $IIA_E$ theory should also gain an extra dimension, and supersymmetry fixes this to be timelike.
The Euclidean $IIA_E$ theory can be quantized using one dimension ($x^9$ say) as a Euclidean
time, with corresponding \lq hatted'  charges that are conserved with respect to the Euclidean
time. At strong coupling, this should match with M-theory on a timelike circle, in a Euclidean time quantization using $x^9$ as the canonical variable.

\section{Discussion}

In this paper we have conjectured that the maximally supersymmetric Euclidean 5 dimensional   Yang-Mills theory (with UV completion) should be identified with the (2,0) theory compactified on a timelike circle of radius
\begin{align}
R = \frac{g^2}{4\pi^2}\ .
\end{align}
We have provided evidence for this by
 examining its superalgebra and matching  BPS solutions in 5D with BPS states in 5+1 dimensions, including states that arise from Lorentz and conformal transformations of the 6D theory. We also argued that some  of the 5D classical solutions we found can also be identified with states of the (2,0) theory with non-compact time but compactified on a spacelike circle.

\subsection{Quantizations}

A given classical theory can admit different inequivalent quantizations. In the case of a Euclidean field theory, there are a number of approaches to quantization. We consider some of them now, and discuss  their relationship to one another and their implications for 5D SYM.

1) There is a quantization where we choose a Euclidean \lq time' ($\tau$) and follow
the canonical quantization for this choice of time. In this approach, there are $\tau$-conserved hatted
charges for configurations falling off in the directions transverse to this choice of \lq time',   and  in this approach we have evidence that there is a compact time emerging for 5D SYM.

In principle there is a different quantum theory for each choice of $\tau$, 
with each focussing on configurations with boundary conditions aligned to that choice of $\tau $.
However, all
give the same picture and the same radius for the extra dimension.
In standard canonical quantization in 4+1, SO(4,1) is broken to SO(4) by
choosing a time and a corresponding set of boundary conditions, and then it is non-trivial to show that the quantum
theory is in fact SO(4,1) Lorentz covariant.
For 5D SYM in 5+0, choosing a Euclidean \lq time' ($\tau$) breaks SO(5) to SO(4). A
key question is then whether the resulting quantum theory is in fact SO(5) covariant. A related question is whether this approach can incorporate states that do not obey the boundary conditions adapted to $\tau$ considered earlier.


2) Path integral quantization as conventionally defined involves a time-slicing of paths in Minkowski space, or a slicing with respect to a Euclidean time for a
Euclidean path integral. It is this slicing that gives rise to the canonical commutation relations at fixed time or
 Euclidean time, recovering the canonical approach.
 
3) Formally, the path integral might be thought of as  generating Euclidean space correlation functions,  calculated 
in  perturbation theory in terms of Euclidean space Feynman
diagrams. The perturbative quantum theory might be thought of as being defined by these correlation functions, which will be SO(5) covariant and are  independent of any choice of Euclidean time.
  As discussed earlier, this is defined by analytically continuing 
$\phi^I \to -i \phi^I$ to give a positive bosonic Euclidean  action, the same as that used for quantizing 4+1
SYM. The same path integral would then have different continuations back to results for 4+1 SYM and for 5+0 SYM.
The path integral appears to involve no explicit choice of Euclidean time (although there is a possibility that this might enter in the continuation back to the real section).
For the 4+1 theory, we expect an extra dimension to emerge at strong coupling, and the correlation functions from the Euclidean path integral should be consistent with this.
For the 5+0 theory, we have the same Euclidean correlation functions and so these too should be consistent with the emergence of an extra dimension. On continuation back to the real section, we expect to find a theory in 5+1 dimensions in both cases.
 The results of this approach seem to be consistent with the Euclidean time canonical quantization. 
 
 Note that a similar path integral emerges in considering the 5+1 (2,0) theory at finite temperature.
 In that case one considers the Euclidean theory on $\R^5\times S^1$, with a thermal circle with radius given by the inverse  temperature in the usual way.
 The standard partition function involves anti-periodic fermions which break supersymmetry, but the insertion of $(-1)^F$ gives an index-like quantity.
Such path integrals with $\R^5$ replaced by e.g. an $S^5$ have been considered in \cite{Kim:2012ava,Kallen:2012zn,Kim:2013nva}.

4) Another approach is to dimensionally reduce a quantum theory in 5+1 dimensions on a time dimension. 
One starts from the conventional canonical quantization in 5+1 dimensions  with real time
and physical charges. This involves choosing a time direction, breaking  
Lorentz symmetry down to $SO(5)$.
One focusses on the subsector of time-independent operators   and uses the
commutation relations from the 5+1 canonical approach. This gives a 5-D
Euclidean quantum theory.
In this picture, states in 5D are naturally associated with time-independent
states or states that have been smeared over time. For example
a static 1-brane  in 5+1 dimensions is time-independent and reduces to an E1-brane in 5-D.
Massless states moving in the $x^5$ direction in $5+1$ dimensions are represented by fields depending on $t-x^5$ but not $t+ x^5$. Smearing over time gives fields independent of $t$ and also independent of $x^5$, and reducing to 5D gives an E1-brane in the $x^5$ direction.
This could be applied to the 5+1 dimensional field theory in section 7, and the     charges  $\tilde \Pi$ etc seem to play a natural role here. This could equally be applied in principle to the (2,0) M5 brane 
theory, and the corresponding   charges $  \Pi_\mu$ appear to be closely related to the $\tilde \Pi_\mu$  arising in section 7.
 
 5) A related approach is to regard the Euclidean 5D theory as the theory in 5+1 dimensions of section 7 with a spacelike vector $C$, and to use a conventional canonical quantisation in 5+1 dimensions. Then 
 Euclidean correlations arising from the 5D SYM  are interpreted as  giving 6D correlation functions at fixed energy and momenta specified by $\tilde \Pi_\mu$.

\subsection{Other Variants of the Theory}

 As discussed in section 1, there are 2 Euclidean SYM theories in 5 dimensions. We have proposed that the one with $SO(5)$ R-symmetry is to be identified with the (2,0) theory on a timelike circle. This, however, leaves the question of what the strong coupling limit of the SYM with $SO(4,1)$ R-symmetry might be. There is in fact another supersymmetric (2,0) theory in 5+1 dimensions, but with R-symmetry $SO(4,1)$ instead of $SO(5)$  \cite{Hull:1999mt}, and this compactified on a timelike circle is the natural strong coupling dual. Note that this theory in 5+1 dimensions is non-unitary, due to the non-compact R-symmetry.
This and related dualities  follow from the dualities and relations found in \cite{Hull:1998vg, Hull:1998ym,Hull:1998fh,Hull:1999mt}.
In 5 dimensions, there are SYM theories in signature 4+1 with R-symmetry SO(5) or SO(4,1), in signature 5+0 with R-symmetry $SO(5)$ or $SO(4,1)$ and in signature 3+2 with R-symmetry $SO(3,2)$.
In 6 dimensions, there are theories with (2,0) symmetry in signature 5+1 with R-symmetry $SO(5)$ or $SO(4,1)$, and in signature 3+3 with R-symmetry $SO(3,2)$  \cite{Hull:1999mt}. The origins of these as world-volume theories of branes are given in  \cite{Hull:1999mt}, and these immediately give the formal strong coupling limits of the various SYM theories.
The  SYM theory in signature 4+1 with R-symmetry   SO(4,1) is dual to the  (2,0) theory in signature 5+1 with R-symmetry   $SO(4,1)$ compactified on a space like circle, while
SYM in signature 3+2 with R-symmetry $SO(3,2)$ is dual to the (2,0) theory in 3+3 dimensions, compactified on a circle.

\section*{Acknowledgements}

CH  is supported  by STFC grant ST/L00044X/1 and EPSRC grant EP/K034456/1. N.L. is supported in part by STFC grant ST/J002798/1. 
Both C.H. and N.L. would like to thank the Corfu Summer Institute for its hospitality.  CH would like to thank CERN for   hospitality.



\section*{Appendix}
\setcounter{equation}{0}

\renewcommand{\theequation}{A.\arabic{equation}}

In this appendix we give some additional details on the singular BPS solutions discussed in section 5.2. Since the solutions are abelian we may first consider the case of a single M5-brane. The solution takes the form \cite{Howe:1997ue}
\begin{equation}
H_{05i} = \pm\partial_i \Phi^6\qquad H_{ijk} = \mp\varepsilon_{05ijkl}\partial^l\Phi^6\qquad \Phi^6 = \langle\Phi^6\rangle - \frac{ Q_E}{4\pi^2 r^2} \ ,
\end{equation}
where $Q_E$ is required to be an integer. 
In terms of the fields of 5D SYM ($\phi^I = g^2\Phi^I$, $F_{ij} = g^2 H_{0ij}$) this solution is 
\begin{equation}
F_{5i} = \pm\partial_i \phi^6\qquad \phi^6 = \langle \phi^6\rangle - \frac{g^2 Q_E}{4\pi^2 r^2} \ ,
\end{equation}
in agreement with (\ref{sdssol}).
We may evaluate the energy per unit length along $x^5$ using the abelian energy-momentum tensor (\ref{Ta}):
\begin{align}
P^ 0 & = \int d^4x T^{(abelian)}_{00}\nonumber\\
&= 2\pi  \int d^4x \frac{ Q_E^2}{(2\pi^2)^2 r^6}\nonumber\\
&= \frac{Q_E^2}{\pi }\int^\infty_\epsilon \frac{dr}{r^3}\nonumber\\
& = \frac{Q_E^2}{2 \pi \epsilon^2 }\ .
\end{align}
Here, as in \cite{Callan:1997kz}, we have introduced a cut-off at $r=\epsilon$ to regulate the divergence. We can physically interpret the cut-off by noting that 
\begin{equation}
\langle \Phi^6\rangle - \Phi^6(\epsilon) =  \frac{1 }{4\pi^2 \epsilon^2}Q_E\ ,
\end{equation}
and hence
\begin{align}
P^0 & =  2\pi   Q_E(\langle \Phi^6\rangle - \Phi^6(\epsilon))\ .
\end{align}

To proceed, we need to find the normalization  of the six dimensional scalar fields $\Phi^I$ in terms of the physical coordinates $x^I$ of 11-dimensional spacetime that are transverse to the M5-brane ($I,J=6,7,8,9,10$). In particular, while the $\Phi^I$ have mass dimension two   the coordinates $x^I$ have mass dimension minus one. To this end, let us compare the scalar contribution to the energy-momentum tensor (\ref{Ta}) with that which arises from a Nambu-Goto action for the M5-brane
\begin{align}
S_{NG} &= -T_{M5} \int d^6x \sqrt{-\det(\eta_{\mu\nu}+\partial_\mu X^I\partial_\nu X^I)}\nonumber\\ &= T_{M5}\int d^6x \left(-1+ \frac{1}{2}\partial_\mu X^I\partial^\mu X^I+\ldots \right)\ ,
\end{align}
where   $X^I$ are the transverse embedding coordinates of the M5-brane in static gauge ({\it i.e.} $X^\mu = x^\mu$, $\mu=0,1,2,...,5$).
Comparing with (\ref{Ta})  we see that $\Phi^I = \sqrt{T_{M5}/2\pi}X^I$. Since    $T_{M5}= \frac{1}{2\pi}(T_{M2})^2 $ we find  $\Phi^I = T_{M2}X^I/2\pi $ and hence the mass per unit length can be written as 
\begin{align}
P^ 0 & =  T_{M2}Q_E(\langle X^6\rangle - X^6(\epsilon))\ .
\end{align}
For a single self-dual string we have $Q_E=1$ and  
hence the divergent energy per unit length is in agreement with the area of a semi-infinite M2-brane that stretches from the M5-brane located at $ x^6  =\langle X^6\rangle $ all the way  out to $x^6 = X^6(\epsilon)\to\infty$. This provides an alternative justification to \cite{Maldacena:1997de} for the $2\pi$ in the $(2,0)$ energy momentum tensor (\ref{Ta}).

Let us now look at the related  solutions (\ref{sdssol}) of 5D  SYM  and interpret them following section 7 as static states of the (2,0) theory on a spacelike circle of radius $R$ but with non-compact time. Since these are static solutions, the energy obtained from (\ref{ThetaTis}) will diverge due to the integral over time. To regulate this we can also make time periodic   with period $2\pi R'$. In this case one finds from (\ref{ThetaTis}) that, for static solutions, 
\begin{align}
P^0 &= \int d^5x T^{00}\nonumber\\
&= \frac{2\pi R}{2\pi R'} \int d^5x \tilde \Theta^{00}
\ .  
\end{align}
According to the discussion at the end of section 7 compactifying time corresponds to making $x^5$ periodic in 5D SYM  with period $2\pi R'$.
Evaluating the  integral now leads to
\begin{equation}
P^0 = 2\pi R  T_{M2}{\rm tr}(Q_E(\langle X^6\rangle - X^6(\epsilon)))\ .
\end{equation}
Just as in the abelian case discussed above this is in agreement with that of  semi-infinite strings that have been wrapped on a spatial circle of radius $R$.

\bibliographystyle{utphys}
\bibliography{time}

\end{document}